\documentclass[a4paper,fleqn,usenatbib]{mnras}

\usepackage{amsmath,amssymb}
\usepackage{mathptmx}
\usepackage{txfonts}

\usepackage[T1]{fontenc}
\usepackage{ae,aecompl}

\usepackage{graphicx}	
\usepackage{amsmath}	
\usepackage{amssymb}	

\def\rpd{\hbox{rad\,d$^{-1}$}}

\def\chisq{\hbox{$\chi^2$}}
\def\chisqr{\hbox{$\chi^2_{\rm r}$}}
\def\msun{\hbox{${\rm M}_{\odot}$}}
\def\mjup{\hbox{${\rm M}_{\rm Jup}$}}

\def\rsun{\hbox{${\rm R}_{\odot}$}}
\def\lsun{\hbox{${\rm L}_{\odot}$}}

\def\mstar{\hbox{$M_{\star}$}}
\def\rstar{\hbox{$R_{\star}$}}
\def\lstar{\hbox{$L_{\star}$}}
\def\teff{\hbox{$T_{\rm eff}$}}
\def\logg{\hbox{$\log g$}}
\def\sn{\hbox{S/N}}
\def\vrad{\hbox{$v_{\rm rad}$}}

\def\kms{\hbox{km\,s$^{-1}$}}
\def\vsini{\hbox{$v \sin i$}}

\def\V{\hbox{${\rm V}$}}
\def\BmV{\hbox{${\rm B-V}$}}
\def\VmRj{\hbox{${\rm V-R_{\rm J}}$}}
\def\AV{\hbox{$A_{\rm V}$}}

\def\em{\it}

\def\degr{\hbox{$^\circ$}}

\def\omeq{\hbox{$\Omega_{\rm eq}$}}
\def\dom{\hbox{$d\Omega$}}
\def\Porb{\hbox{$P_{\rm orb}$}}
\def\Prot{\hbox{$P_{\rm rot}$}}

\newcommand{\caii}{\hbox{Ca$\;${\sc ii}}}
\newcommand{\fei}{\hbox{Fe$\;${\sc i}}}

\newcommand{\hei}{\hbox{He$\;${\sc i}}}

\title[Magnetometry \& velocimetry of the wTTSs V819~Tau \& V830~Tau]{Magnetic activity and hot Jupiters of young Suns: \\ the weak-line T~Tauri stars V819~Tau and V830~Tau}

\author[J.-F.~Donati et al.]{ 
J.-F.~Donati$^{1,2}$\thanks{E-mail: jean-francois.donati@irap.omp.eu },
E.~H\'ebrard$^{1,2}$, G.A.J.~Hussain$^{3,1}$, C.~Moutou$^4$, L.~Malo$^4$, K.~Grankin$^5$, 
\newauthor 
A.A.~Vidotto$^6$, S.H.P.~Alencar$^7$, S.G.~Gregory$^8$, M.M.~Jardine$^8$, G.~Herczeg$^9$, 
\newauthor 
J.~Morin$^{10}$, R.~Fares$^{11}$, F.~M\'enard$^{12}$, J.~Bouvier$^{13,14}$, X.~Delfosse$^{13,14}$, R.~Doyon$^{15}$, 
\newauthor 
M.~Takami$^{16}$, P.~Figueira$^{17}$, P.~Petit$^{1,2}$, I.~Boisse$^{18,19}$ and the MaTYSSE collaboration 
\\
$^1$ Universit\'e de Toulouse, UPS-OMP, IRAP, 14 avenue E.~Belin, Toulouse, F--31400 France \\
$^2$ CNRS, IRAP / UMR 5277, Toulouse, 14 avenue E.~Belin, F--31400 France \\
$^3$ ESO, Karl-Schwarzschild-Str.\ 2, D-85748 Garching, Germany \\
$^4$ CFHT Corporation, 65-1238 Mamalahoa Hwy, Kamuela, Hawaii 96743, USA \\
$^5$ Crimean Astrophysical Observatory, Nauchny, Crimea 298409 \\
$^6$ Observatoire de Gen\`eve, Chemin des Maillettes 51, CH-1290 Versoix, Switzerland \\
$^7$ Departamento de F\`{\i}sica -- ICEx -- UFMG, Av. Ant\^onio Carlos, 6627, 30270-901 Belo Horizonte, MG, Brazil \\
$^8$ SUPA, School of Physics and Astronomy, Univ.\ of St~Andrews, St~Andrews, Scotland KY16 9SS, UK \\
$^9$ Kavli Institute for Astronomy and Astrophysics, Peking University, Yi He Yuan Lu 5, Haidian Qu, Beijing 100871, China \\
$^{10}$ LUPM, Universit\'e de Montpellier, CNRS, place E.~Bataillon, F--34095 Montpellier, France \\
$^{11}$ INAF - Osservatorio Astrofisico di Catania, via S. Sofia 78, I--95123 Catania, Italy \\
$^{12}$ CNRS, UMI-FCA / UMI 3386, France, and Universidad de Chile, Santiago, Chile  \\
$^{13}$ Universit\'e Grenoble Alpes, IPAG, BP~53, F--38041 Grenoble C\'edex 09, France \\
$^{14}$ CNRS, IPAG / UMR 5274,  BP~53, F--38041 Grenoble C\'edex 09, France \\
$^{15}$ D\'epartement de physique, Universit\'e de Montr\'eal, C.P.~6128, Succursale Centre-Ville, Montr\'eal, QC, Canada  H3C 3J7 \\
$^{16}$ Institute of Astronomy and Astrophysics, Academia Sinica, PO Box 23-141, 106, Taipei, Taiwan \\
$^{17}$ Centro de Astrof\`\i sica, Universidade do Porto, Rua das Estrelas, 4150-762 Porto, Portugal \\
$^{18}$ Universit\'e Aix-Marseille, LAM, F--13388 Marseille, France \\ 
$^{19}$ CNRS, LAM / UMR~7326, F--13388 Marseille, France 
}

\date{Accepted 2015 August 6.  Received 2015 August 5; in original form 2015 May 8} 

\pubyear{2015}

\begin{document}
\label{firstpage}
\pagerange{\pageref{firstpage}--\pageref{lastpage}}
\maketitle

\begin{abstract}
We report results of a spectropolarimetric and photometric monitoring of the weak-line T~Tauri stars (wTTSs) V819~Tau and V830~Tau
within the MaTYSSE programme, involving the ESPaDOnS spectropolarimeter at the Canada-France-Hawaii Telescope.
At $\simeq$3~Myr, both stars dissipated their discs recently and are interesting objects for probing star and planet formation.
Profile distortions and Zeeman signatures are detected in the unpolarized and circularly-polarized lines, 
whose rotational modulation we modelled using tomographic imaging, yielding brightness and magnetic maps for both stars.

We find that the large-scale magnetic fields of V819~Tau and V830~Tau are mostly poloidal and can be approximated
at large radii by 350--400~G dipoles tilted at $\simeq$30\degr\ to the rotation axis.  They are significantly weaker than the field of 
GQ~Lup, an accreting classical T~Tauri star (cTTS) with similar mass and age which can be used to compare the magnetic properties of 
wTTSs and cTTSs.  The reconstructed brightness maps of both stars include cool spots and warm plages.  Surface differential rotation is small, 
typically $\simeq$4.4$\times$ smaller than on the Sun, in agreement with previous results on wTTSs. 

Using our Doppler images to model the activity jitter and filter it out from the radial velocity (RV) curves, we obtain RV residuals
with dispersions of 0.033 and 0.104~\kms\ for V819~Tau and V830~Tau respectively.  RV residuals suggest that a hot
Jupiter may be orbiting V830~Tau, though additional data are needed to confirm this preliminary result.  We find no evidence 
for close-in giant planet around V819~Tau.
\end{abstract}




\section{Introduction}
\label{sec:int}

In the last decades, our understanding of the key role that magnetic fields play during the first stages of the life of stars
and their planets improved significantly, through both theoretical and observational advances \citep[e.g.,][]{Andre09}.
This is true in particular for low-mass pre-main-sequence (PMS) stars in the T~Tauri star (TTS) phase, aged 0.5--10~Myr,
that recently emerged from their dust cocoons and are still in a phase of gravitational contraction towards the main sequence (MS).
Being either classical T-Tauri stars (cTTSs) when still surrounded by and accreting from massive, planet-forming accretion discs, or weak-line
T-Tauri stars (wTTSs) once they have exhausted the gas in their inner discs, TTSs have been the subject of intense scrutiny at all wavelengths
in recent decades given their interest for benchmarking the scenarios currently invoked to explain low-mass star and planet
formation.

Magnetic fields of cTTSs are known to influence and sometimes even trigger several of the main physical processes playing a role
at this stage, like accretion, outflows and angular momentum transport.  As a result, they largely control the rotational
evolution of low-mass PMS stars \citep[e.g.,][]{Bouvier07, Frank14}.  More specifically, large-scale fields of cTTSs can
evacuate the central regions of accretion discs, funnel the disc material onto the stars, and enforce corotation between
cTTSs and their inner-disc Keplerian flows.  This is causing cTTSs to rotate much more slowly than expected from the contraction
and accretion of the high-angular-momentum disc material \citep[e.g.,][]{Davies14}.  Fields of cTTSs and of their discs can also
affect the formation and migration of planets \citep[e.g.,][]{Baruteau14}.  Last but not least, magnetic fields of cTTSs / wTTSs are known
to trigger thermally-driven winds through the heating provided by accretion shocks and / or Alfv\'en waves \citep[e.g.,][]{Cranmer09,
Cranmer11}, leading to flares, coronal-mass ejections and angular momentum loss \citep[e.g.,][]{Matt12, Aarnio12}.

For these reasons, characterizing magnetic fields of cTTSs and wTTSs through observations is a crucial step to guide
theoretical models towards more physical realism and reliable predictions.  Although first detected more than 15~yrs ago
\citep[e.g.,][]{Johns99b, Johns07}, magnetic fields of TTSs are not yet fully characterized.  In particular, field
topologies of cTTSs have only recently been unveiled for a dozen stars \citep[e.g.,][]{Donati07, Hussain09, Donati10b, Donati13}
through the MaPP (Magnetic Protostars and Planets) Large Observing Programme allocated on the 3.6~m Canada-France-Hawaii
Telescope (CFHT) with the ESPaDOnS high-resolution spectropolarimeter (550~hr of clear time over semester 2008b to 2012b).
This first survey revealed for instance that large-scale fields of cTTSs remain relatively simple and mainly poloidal when
the host star is still fully or largely convective, but become much more complex when the star turns mostly radiative
\citep{Gregory12, Donati13}.  This survey also demonstrated that these fields vary over a few years \citep[e.g.,][]{Donati11,
Donati12, Donati13} and resemble those of mature stars with comparable internal structures \citep{Morin08b}, hinting at a
dynamo rather than fossil origin.

The situation is even worse for wTTSs, with only two of them magnetically imaged to date \citep[namely V410~Tau and LkCa~4,][]{Skelly10, Donati14}.
In both cases, the large-scale fields found are unexpected with respect to what we already know on cTTSs and MS dwarfs with
similar internal structures.  The most recent observations, focussed on the wTTS LkCa~4, revealed the presence of a ring of intense
toroidal field encircling the star at low latitudes, despite LkCa~4 being most likely fully convective and thus not supposed to exhibit
this kind of magnetic feature \citep{Donati14}.  Clearly, a more systematic exploration of large-scale fields of wTTSs needs to be
carried out in a similar manner to that of cTTSs.  This is the goal of the new MaTYSSE (Magnetic Topologies of
Young Stars and the Survival of close-in giant Exoplanets) Large Programme, allocated at CFHT over semesters 2013a-2016b (510~hr)
with complementary observations with the NARVAL spectropolarimeter on the 2-m T\'elescope Bernard Lyot at Pic du Midi in France 
(420~hr, allocated) and with the HARPS spectropolarimeter at the 3.6-m ESO Telescope at La Silla in Chile (70~hr, allocated).  
MaTYSSE also allows us to study magnetic winds of wTTSs and the corresponding spin-down rates \citep[e.g.,][]{Vidotto09, Vidotto10, Vidotto14};  
it also offers a novel option for assessing with improved sensitivity the potential presence of hot Jupiters (hJs) at an early stage in the 
life of low-mass stars and their planets, and thus for verifying whether core accretion coupled to migration is indeed the best option for 
forming close-in giant planets.  

In practice, MaTYSSE aims at carrying out a survey of about 30~wTTSs, along with a long-term monitoring of $\simeq$5~cTTSs for which 
temporal variability of the large-scale magnetic field was already detected with MaPP.  Rather than publishing the whole MaTYSSE data set as a single 
study (which would imply both long delays to collect the data and carry out the analysis, and an extremely long paper), our publication plan is to 
slice the work into smaller steps, by dedicating each forthcoming paper to only a few stars with similar properties.  This ensures in particular that 
a significant fraction of observations can be published quickly and before data become public, and to advertise the MaTYSSE results so that additional 
observing time (e.g., on HARPS-Pol) can be obtained in the meantime through complementary programmes.  The statistical properties of the whole sample 
will be presented and discussed in a final wrap-up paper.   

Following our initial MaTYSSE paper \citep{Donati14}, we present here a full analysis of the phase-resolved spectropolarimetric data sets collected
on two new wTTSs, namely V819~Tau and V830~Tau.  Both are well studied single wTTSs \citep[from direct imaging and spectroscopic monitoring, e.g.,][]{Kraus11, Nguyen12} 
and bona fide members of the Taurus L1495 dark cloud \citep{Xiao12}, showing clear photometric variations of large amplitudes \citep[][]{Grankin08, Xiao12}.  
V819~Tau is also known to exhibit a weak infrared excess indicating gas-poor remnants of a primordial disc or the early presence a 
warm debris disc \citep{Furlan09b, Cieza13}.

We start our paper by documenting our
observations (Sec.~\ref{sec:obs}) and reviewing the spectral characteristics and evolutionary status of both stars (Sec.~\ref{sec:evo});
we then describe the results we obtain by applying our tomographic modelling technique to these new data (Sec.~\ref{sec:mod}), use
them to filter out the activity jitter from RV curves and look for the potential presence of hJs around both stars (Sec.~\ref{sec:fil}).
Finally, we summarize our main results and discuss their implications for our understanding of low-mass star and planet formation
(Sec.~\ref{sec:dis}).

\section{Observations}
\label{sec:obs}

V819~Tau and V830~Tau were both observed in 2014~December and 2015~January using ESPaDOnS at the CFHT.
ESPaDOnS collects stellar spectra spanning the entire optical domain (from 370 to 1,000~nm) at a resolving power of 65,000
(i.e., a resolved velocity element of 4.6~\kms) over the full wavelength range \citep{Donati03}.
A total of 15 circularly-polarized (Stokes $V$) and unpolarized (Stokes $I$) spectra were collected for both stars over a timespan of 28 nights,
corresponding to about 5 rotation cycles for V819~Tau and 10 cycles for V830~Tau.  Time sampling was irregular in the first two thirds of the
run (as a result of poor weather) but denser in the last third, yielding a reasonably good phase coverage for both stars
despite a few remaining phase gaps of 0.15--0.20~cycle.

All polarisation spectra consist of 4 individual subexposures
taken in different polarimeter configurations to allow the removal of all spurious polarisation signatures at first order.
All raw frames are processed as described in the previous papers of the series
\citep[e.g.,][]{Donati10b, Donati11, Donati14}, to which the reader is referred for more information.
The peak signal-to-noise ratios (\sn, per 2.6~\kms\ velocity bin) achieved on the
collected spectra range between 140 and 220 (median 190) for V819~Tau, and between 140 and 180 (median 170) for V830~Tau,
depending mostly on weather/seeing conditions.
All spectra are automatically corrected of spectral shifts resulting from instrumental effects (e.g., mechanical flexures, temperature or pressure variations)
using telluric lines as a reference.  Though not perfect, this procedure provides spectra with a relative RV precision of better than 0.030~\kms\
\citep[e.g.,][]{Moutou07, Donati08b}.  The full journal of observations is presented in Table~\ref{tab:log} for both stars.

\begin{table}
\caption[]{Journal of ESPaDOnS observations of V819~Tau (first 15 lines) and V830~Tau (last 15 lines) collected from
mid 2014~December to mid 2015~January.  Each observation consists of a sequence of 4 subexposures, each lasting 1400~s
and 700~s for V819~Tau and V830~Tau respectively (except on 2015 Jan~15 for which subexposures on V819~Tau were shortened to 900~s).
Columns $1-4$ respectively list (i)~the UT date of the observation, (ii) the
corresponding UT time (at mid-exposure), (iii)~the Barycentric Julian Date (BJD), 
and (iv)~the peak signal to noise ratio (per 2.6~\kms\ velocity bin)
of each observation.  Column 5 lists the rms noise level (relative to the unpolarized continuum level
$I_{\rm c}$ and per 1.8~\kms\ velocity bin) in the circular polarization profile
produced by Least-Squares Deconvolution (LSD), while column~6 indicates the
rotational cycle associated with each exposure (using the ephemerides given by
Eq.~\ref{eq:eph}).  }
\begin{tabular}{cccccc}
\hline
Date   & UT      & BJD      & \sn\ & $\sigma_{\rm LSD}$ & Cycle \\
(2014) & (hh:mm:ss) & (2,457,000+) &      &   (0.01\%)  &   \\
\hline
Dec 19 & 12:07:54 & 11.01081 & 140 & 3.9 & 0.002 \\
Dec 20 & 07:55:28 & 11.83548 & 210 & 2.5 & 0.151 \\
Dec 21 & 07:17:08 & 12.80882 & 210 & 2.4 & 0.327 \\
Dec 22 & 08:14:51 & 13.84886 & 200 & 2.6 & 0.515 \\
Dec 28 & 10:50:17 & 19.95655 & 190 & 2.8 & 1.619 \\
Dec 29 & 07:37:39 & 20.82273 & 180 & 2.9 & 1.776 \\
Dec 30 & 06:10:02 & 21.76183 & 200 & 2.6 & 1.946 \\
Jan 07 & 07:19:42 & 29.80976 & 190 & 2.7 & 3.401 \\
Jan 08 & 06:20:11 & 30.76836 & 190 & 2.7 & 3.574 \\
Jan 09 & 06:15:12 & 31.76484 & 190 & 2.9 & 3.754 \\
Jan 10 & 06:15:60 & 32.76533 & 170 & 3.2 & 3.935 \\
Jan 11 & 07:25:26 & 33.81348 & 220 & 2.3 & 4.125 \\
Jan 12 & 06:28:22 & 34.77379 & 200 & 2.4 & 4.298 \\
Jan 14 & 06:17:48 & 36.76631 & 190 & 2.8 & 4.658 \\
Jan 15 & 06:12:46 & 37.76274 & 150 & 3.5 & 4.839 \\
\hline
Dec 20 & 09:13:43 & 11.88992 & 170 & 2.8 & 0.033 \\
Dec 21 & 08:33:48 & 12.86217 & 170 & 2.8 & 0.387 \\
Dec 22 & 09:29:48 & 13.90104 & 180 & 2.7 & 0.766 \\
Dec 28 & 12:19:59 & 20.01897 & 140 & 3.6 & 2.999 \\
Dec 29 & 08:53:59 & 20.87588 & 160 & 3.0 & 3.311 \\
Dec 30 & 07:26:53 & 21.81535 & 160 & 2.9 & 3.654 \\
Jan 07 & 08:35:55 & 29.86285 & 170 & 2.8 & 6.590 \\
Jan 08 & 07:36:48 & 30.82174 & 180 & 2.7 & 6.940 \\
Jan 09 & 07:31:38 & 31.81810 & 170 & 2.9 & 7.303 \\
Jan 10 & 07:32:25 & 32.81858 & 150 & 3.3 & 7.668 \\
Jan 11 & 08:42:02 & 33.86686 & 180 & 2.6 & 8.051 \\
Jan 12 & 05:12:50 & 34.72151 & 170 & 2.9 & 8.362 \\
Jan 13 & 05:03:32 & 35.71500 & 160 & 3.0 & 8.725 \\
Jan 14 & 05:02:24 & 36.71414 & 170 & 2.8 & 9.089 \\
Jan 15 & 07:12:21 & 37.80431 & 170 & 2.9 & 9.487 \\
\hline
\end{tabular}
\label{tab:log}
\end{table}

Rotational cycles of V819~Tau and V830~Tau (noted $E_1$ and $E_2$ in the following equation) are computed from Barycentric Julian Dates (BJDs)
according to the ephemerides:
\begin{eqnarray}
\mbox{BJD} \hbox{\rm ~(d)} & = & 2,457,011.0 + 5.53113 E_1  \hbox{\rm ~~(for V819 Tau)} \nonumber\\
\mbox{BJD} \hbox{\rm ~(d)} & = & 2,457,011.8 + 2.74101 E_2  \hbox{\rm ~~(for V830 Tau)}
\label{eq:eph}
\end{eqnarray}
in which the photometrically-determined rotation periods \Prot\ \citep[equal to 5.53113 and 2.74101~d respectively,][]{Grankin13}
are taken from the literature and the initial Julian dates (2,457,011.0 and 2,457,011.8~d) are chosen arbitrarily.

Least-Squares Deconvolution \citep[LSD,][]{Donati97b} was applied to all spectra.
The line list we employed for LSD is computed from an {\sc Atlas9} LTE model atmosphere \citep{Kurucz93}
featuring an effective temperature of $\teff=4,250$~K and a logarithmic gravity (in cgs unit) of $\logg=4.0$, appropriate for both V819~Tau and V830~Tau (see Sec.~\ref{sec:evo}).
As usual, only moderate to strong atomic spectral lines are included in this list \citep[see, e.g.,][for more details]{Donati10b}.
Altogether, about 7,800 spectral features (with about 40\% from \fei) are used in this process.
Expressed in units of the unpolarized continuum level $I_{\rm c}$, the average noise levels of the resulting Stokes $V$
LSD signatures range from 2.3 to 3.9$\times10^{-4}$ per 1.8~\kms\ velocity bin - with a median value of 2.8$\times10^{-4}$ for both stars.

\begin{figure}
\hbox{\includegraphics[scale=0.35,angle=-90]{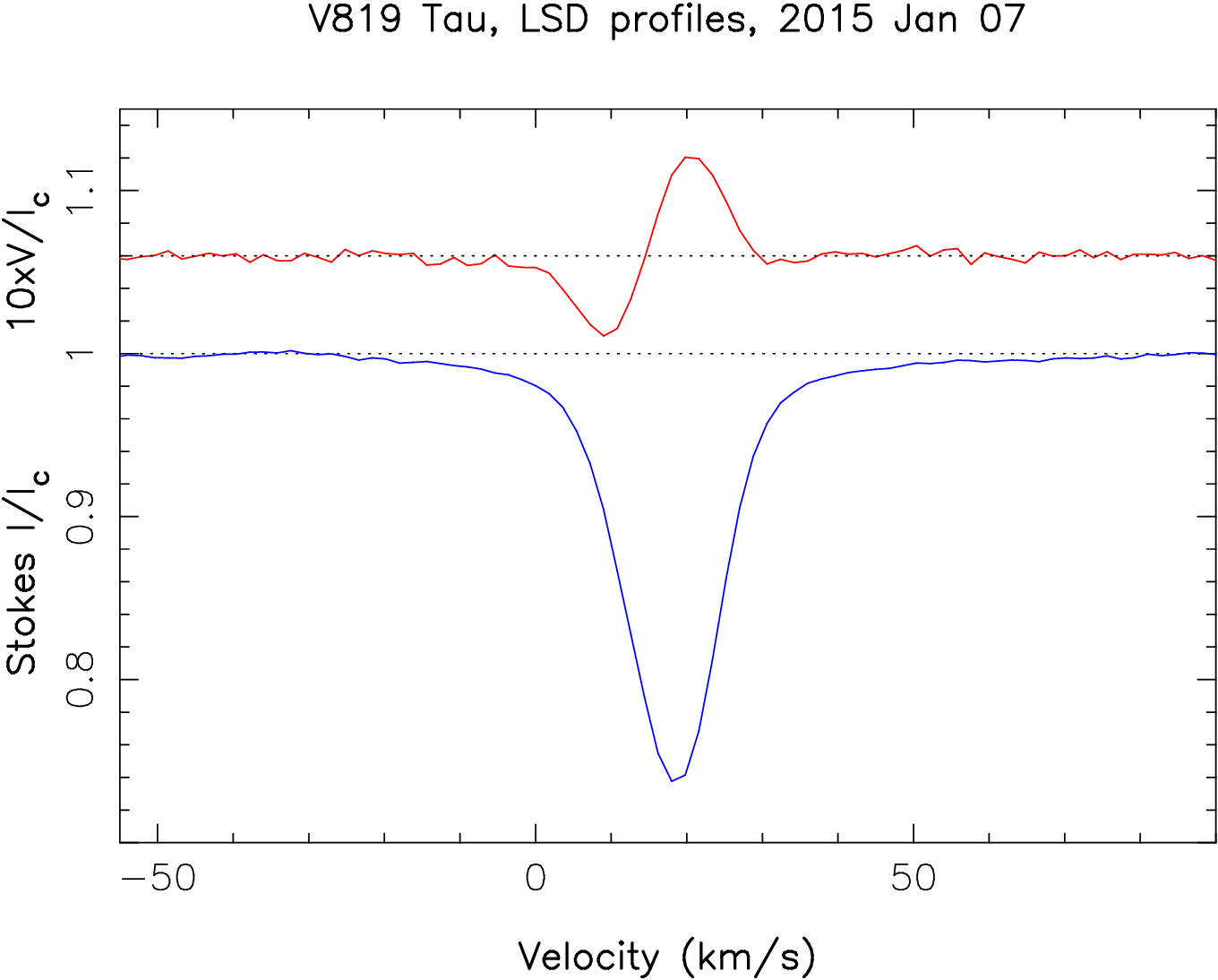} \vspace{3mm}}
\includegraphics[scale=0.35,angle=-90]{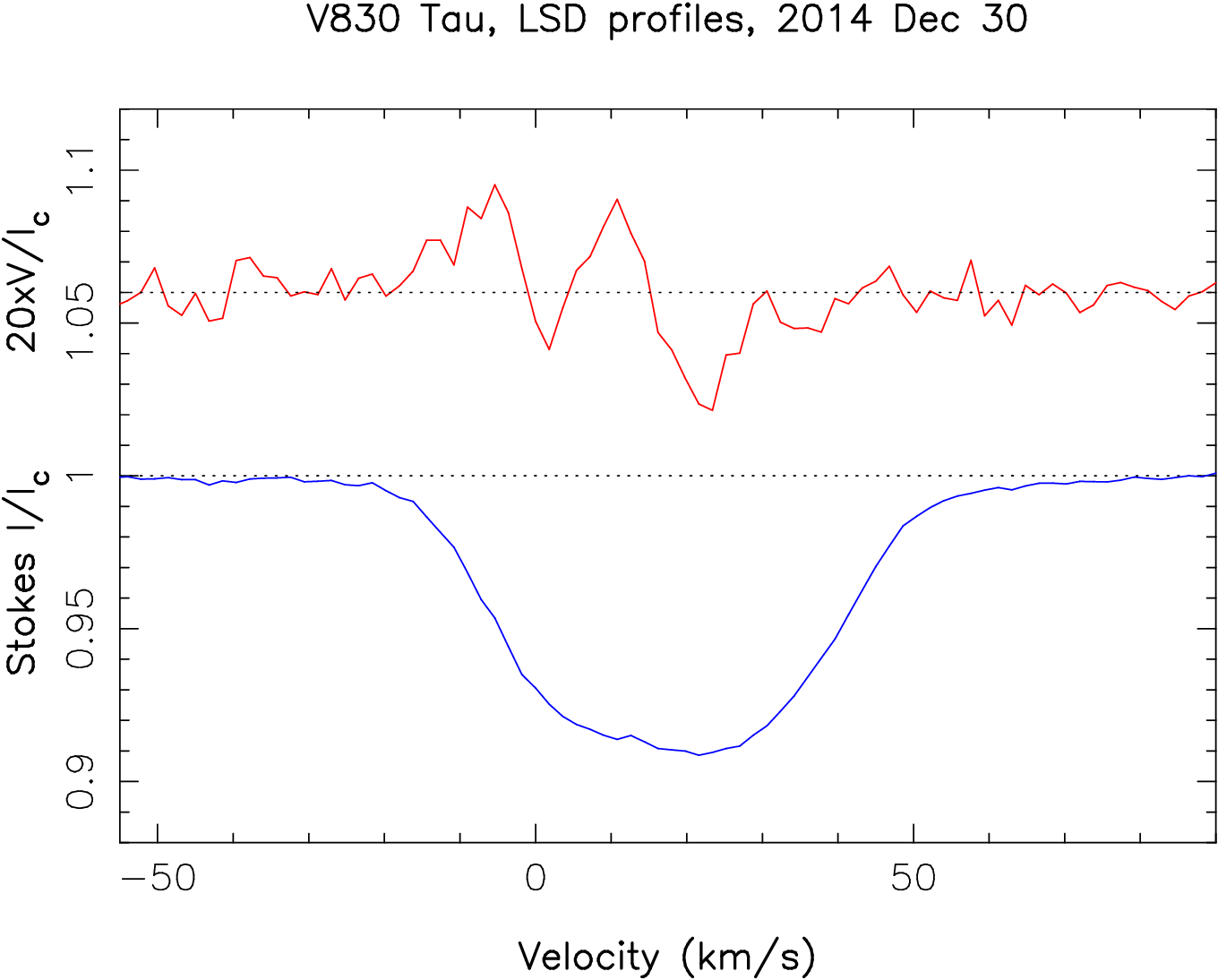}
\caption[]{LSD circularly-polarized (Stokes $V$, top/red curve) and unpolarized (Stokes $I$, bottom/blue curve)
profiles of V819~Tau (top panel) and V830~Tau (bottom panel) collected on 2015~Jan.~07 (cycle 3.401) and
2014~Dec.~30 (cycle 3.654).  Clear Zeeman signatures are detected in the LSD Stokes $V$ profile of both stars
(with a complex shape in the case of V830~Tau), in conjunction with the unpolarized line profiles.
The mean polarization profiles are expanded (by a factor of 10 and 20 for V819~Tau and V830~Tau respectively)
and shifted upwards by 1.06 for display purposes.  }
\label{fig:lsd}
\end{figure}

Zeeman signatures are detected at all times in Stokes $V$ LSD profiles (see Fig.~\ref{fig:lsd} for an example),
featuring amplitudes of 0.3--1\%, i.e., indicative of significant large-scale fields.
Asymmetries and / or distortions are also visible in Stokes $I$ LSD profiles, suggesting the presence of
brightness inhomogeneities at the surfaces of V819~Tau and V830~Tau at the time of our observations.

Contemporaneous BVR$_{\rm J}$ photometric observations were also collected from the Crimean Astrophysical Observatory (CrAO) 1.25~m telescope between 2014~Aug and
Dec for both stars, indicating that V819~Tau and V830~Tau were exhibiting brightness modulations with full amplitudes of 0.45 and
0.10~mag in \V\ (see Table~\ref{tab:pho}) and periods of $5.525\pm0.015$ and $2.741\pm0.005$~d respectively
\citep[compatible within error bars with the average periods of][used to phase our spectroscopic data, see Eq.~\ref{eq:eph}]{Grankin13}.
These photometric variations \citep[unusually large for V819~Tau and small for V830~Tau, see Fig.~12a and Table~3 of][]{Grankin08} are
commonly attributed to the presence of brightness features at the surface of both stars.

\begin{table}
\caption[]{Journal of contemporaneous CrAO multicolour photometric observations of V819~Tau (first 22 lines) and V830~Tau (last 19 lines) collected in late 2014,
respectively listing the UT date and Heliocentric Julian Date (HJD) of the observation, the measured \V\ magnitude,
\BmV\ and \VmRj\ Johnson photometric colours, and the corresponding rotational phase (using again the ephemerides given by
Eq.~\ref{eq:eph}). }
\begin{tabular}{cccccc}
\hline
Date   & HJD      & \V    & \BmV & \VmRj & Phase \\
(2013) & (2,456,000+) & (mag) &      &      &       \\
\hline
Aug 26 & 896.5665 & 12.824 & 1.512 & 1.461 & 0.311 \\
Aug 27 & 897.5395 & 12.954 & 1.488 & 1.478 & 0.487 \\
Aug 29 & 899.5421 & 13.124 & 1.577 & 1.520 & 0.849 \\
Aug 30 & 900.5427 & 12.911 & 1.486 & 1.522 & 0.030 \\
Aug 31 & 901.5438 & 12.788 & 1.522 & 1.487 & 0.211 \\
Sep 01 & 902.5487 & 12.922 & 1.524 & 1.509 & 0.393 \\
Sep 02 & 903.5425 & 13.154 & 1.549 & 1.614 & 0.572 \\
Sep 04 & 905.5024 & 12.965 & 1.558 & 1.518 & 0.927 \\
Sep 05 & 906.5354 & 12.801 & 1.549 & 1.470 & 0.113 \\
Sep 20 & 921.5742 & 13.142 & 1.499 & 1.544 & 0.832 \\
Sep 20 & 921.5742 & 13.204 & 1.519 & 1.522 & 0.832 \\
Oct 05 & 936.6035 & 13.099 & 1.618 & 1.447 & 0.549 \\
Oct 15 & 946.5991 & 12.898 & 1.496 & 1.541 & 0.357 \\
Oct 26 & 957.4906 & 12.838 & 1.514 & 1.485 & 0.326 \\
Oct 28 & 959.5578 & 13.259 & 1.572 & 1.578 & 0.700 \\
Nov 02 & 964.4855 & 13.163 & 1.658 & 1.539 & 0.590 \\
Nov 05 & 967.6033 & 12.784 & 1.656 & 1.437 & 0.154 \\
Nov 05 & 967.6106 & 12.795 & 1.518 & 1.513 & 0.155 \\
Nov 13 & 975.4293 & 13.159 & 1.611 & 1.574 & 0.569 \\
Nov 14 & 976.4257 & 13.230 & 1.558 & 1.592 & 0.749 \\
Dec 13 & 1005.4237 & 12.956 & 1.615 & 1.516 & 0.992 \\
Dec 14 & 1006.5232 & 12.804 & 1.586 & 1.506 & 0.191 \\
\hline
Aug 22 & 892.5484 & 12.321 & 1.351 & 1.319 & 0.494 \\
Aug 27 & 897.5627 & 12.297 & 1.403 & 1.297 & 0.323 \\
Aug 29 & 899.5574 & 12.209 & 1.370 & 1.270 & 0.051 \\
Aug 30 & 900.5540 & 12.288 & 1.358 & 1.311 & 0.414 \\
Aug 31 & 901.5514 & 12.238 & 1.368 & 1.296 & 0.778 \\
Sep 01 & 902.5578 & 12.262 & 1.378 & 1.288 & 0.145 \\
Sep 02 & 903.5502 & 12.321 & 1.375 & 1.351 & 0.507 \\
Sep 04 & 905.5599 & 12.304 & 1.389 & 1.286 & 0.241 \\
Sep 05 & 906.5443 & 12.289 & 1.387 & 1.298 & 0.600 \\
Oct 05 & 936.5812 & 12.278 & 1.398 & 9.999 & 0.558 \\
Oct 15 & 946.5610 & 12.297 & 1.405 & 1.303 & 0.199 \\
Oct 28 & 959.5923 & 12.220 & 1.365 & 1.298 & 0.953 \\
Nov 02 & 964.5193 & 12.274 & 1.387 & 1.275 & 0.751 \\
Nov 05 & 967.6268 & 12.210 & 9.999 & 9.999 & 0.884 \\
Nov 05 & 967.6307 & 12.224 & 1.345 & 9.999 & 0.886 \\
Nov 13 & 975.4453 & 12.280 & 1.379 & 1.294 & 0.737 \\
Nov 14 & 976.4437 & 12.277 & 1.351 & 1.321 & 0.101 \\
Dec 13 & 1005.4898 & 12.319 & 1.367 & 1.326 & 0.698 \\
Dec 14 & 1006.4869 & 12.268 & 1.403 & 1.309 & 0.062 \\
\hline
\end{tabular}
\label{tab:pho}
\end{table}

\section{Evolutionary status of V819~Tau and V830~Tau}
\label{sec:evo}

Applying to our highest \sn\ spectra the automatic spectral classification tool especially developed in the context of MaPP and MaTYSSE,
inspired from that of \citet{Valenti05} and discussed in a previous paper \citep{Donati12}, we find that
the photospheric temperatures and logarithmic gravities of V819~Tau and V830~Tau are respectively equal, for both stars, to
$\teff=4250\pm50$~K and $\logg=3.9\pm0.2$ (with $g$ in cgs units).  Repeating the operation on different spectra of our data set yields consistent 
parameters, with differences between results within the quoted error bars.  The temperature we obtain is higher than that often quoted in the 
literature for both stars \citep[in the range 3900--4100~K, see, e.g.,][]{Sestito08, Furlan09b, Grankin13, Herczeg14}, and derived in most cases from 
either photometry or low-resolution spectroscopy.  In the particular case of V830~Tau, our estimate is in agreement with the only other 
spectroscopic study based on high-resolution data that we know of in the refereed literature \citep{Schiavon95}.  
Our result also confirms that V819~Tau and V830~Tau are warmer than our previous target 
LkCa~4 \citep[whose temperature we estimated to be $\teff=4100\pm50$~K,][]{Donati14}.  
From the difference between the \BmV\ index expected at this temperature \citep[$1.15\pm0.02$, as estimated through an interpolation within Table~6 of 
of][]{Pecaut13} and the averaged value measured 
for both V819~Tau and V830~Tau \citep[equal to $1.55\pm0.05$ and $1.35\pm0.05$ respectively, see][see also 
Table~\ref{tab:pho}]{Kenyon95, Grankin08}, we derive that 
the amounts of visual extinction \AV\ that our two targets suffer are equal to $1.24\pm0.15$ for V819~Tau and $0.62\pm0.15$ for V830~Tau \citep[in reasonable agreement
with other independent results, e.g.,][]{Herczeg14}.  

\begin{figure}
\includegraphics[scale=0.35,angle=-90]{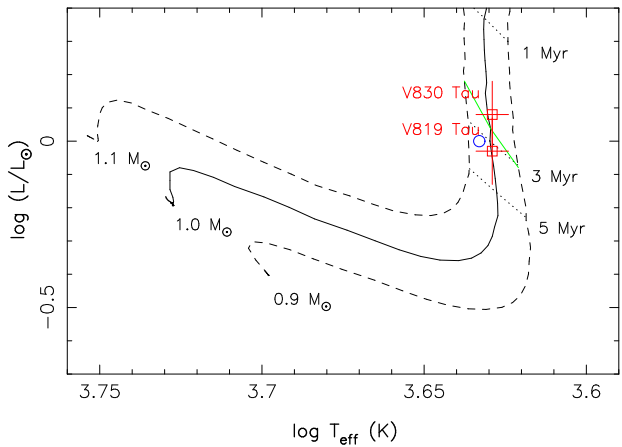}
\caption[]{Observed (open squares and error bars) location of V819~Tau and V830~Tau in the HR diagram.
The PMS evolutionary tracks (full and dashed lines) and corresponding isochrones \citep[dotted lines, all from][]{Siess00} assume solar
metallicity and include convective overshooting.  The blue circle indicates the location of the cTTS GQ~Lup according to \citet{Donati12}.  
The green line depicts where models predict PMS stars start developing their radiative core as they contract towards the main sequence. }
\label{fig:hrd}
\end{figure}

Long-term photometric monitoring of V819~Tau and V830~Tau indicates that their maximum brightnesses correspond to \V\ magnitudes of 12.8 and 11.9 respectively 
\citep{Grankin08};  assuming a spot coverage of $\simeq$25\% of the visible stellar hemisphere at maximum brightness, typical for active stars (and caused by, e.g., 
the presence of high-latitude cool spots 
or by a network of small spots evenly spread in longitude), we derive unspotted \V\ magnitudes of $12.5\pm0.2$ and $11.6\pm0.2$ for V819~Tau and V830~Tau respectively.  
Using the visual bolometric correction expected for the adequate photospheric temperature \citep[equal to $-0.86\pm0.05$,][]{Pecaut13} and the accurate distance 
estimate of the Taurus L1495 dark cloud \citep[$131\pm3$~pc, corresponding to a distance modulus equal to $5.59\pm0.05$,][]{Torres12}, we finally obtain 
bolometric magnitudes of $4.81\pm0.25$ and $4.53\pm0.25$, or equivalently logarithmic luminosities relative to the Sun of $-0.03\pm0.10$ and $0.08\pm0.10$ 
for V819~Tau and V830~Tau respectively.  Coupling with the photospheric temperature obtained previously, we find respective radii of $1.8\pm0.2$ and 
$2.0\pm0.2$~\rsun\ for our two target stars.  

\begin{table}
\caption[]{Main parameters of V819~Tau and V830~Tau as derived from our study except when indicated 
(with G13 and T12 standing for \citealt{Grankin13} and \citealt{Torres12} respectively), with \vrad\ noting 
the RV that the star would have if unspotted (as inferred from the modeling of Sec.~\ref{sec:mod}).  }  
\begin{tabular}{llll}
\hline
                     & V819~Tau       & V830~Tau        & Reference \\
\hline
\mstar\ (\msun)      & $1.00\pm0.05$  & $1.00\pm0.05$   & \\ 
\rstar\ (\rsun)      & $1.8\pm0.2$    & $2.0\pm0.2$     & \\ 
age (Myr)            & $\simeq3.2$    & $\simeq2.2$     & \\ 
\teff\ (K)           & $4250\pm50$    & $4250\pm50$     & \\ 
$\log(\lstar/\lsun)$ & $-0.03\pm0.10$ & $0.08\pm0.10$   & \\ 
\Prot\ (d)           & 5.53113        & 2.74101         & G13 \\ 
\vsini\ (\kms)       & $9.5\pm0.5$    & $30.5\pm0.5$    & \\ 
\vrad\ (\kms)        & $16.6\pm0.1$   & $17.5\pm0.1$    & \\ 
$i$ (\degr)          & $35\pm10$      & $55\pm10$       & \\ 
distance (pc)        & $131\pm3$      & $131\pm3$       & T12 \\ 
\hline
\end{tabular}
\label{tab:par}
\end{table}

The rotation periods of both V819~Tau and V830~Tau are well determined from long-term multi-colour photometric monitoring, and equal to 
5.53113~d and 2.74101~d \citep{Grankin13} respectively;  estimates from different studies \citep[e.g.,][]{Xiao12}  
agree to a precision of better than 1\%, indicative of robust and reliable measurements.  Coupling these rotation periods 
along with our measurements of the line-of-sight-projected equatorial rotation velocities \vsini\ of V819~Tau and V830~Tau 
(respectively equal to $9.5\pm0.5$ and $30.5\pm0.5$~\kms, see Sec.~\ref{sec:mod}), we can infer that the $\rstar \sin i$ values 
of both stars are $1.04\pm0.05$~\rsun\ and $1.65\pm0.03$~\rsun, where \rstar\ and $i$ note the radius of the stars and the inclination 
of their rotation axis to the line of sight.  Comparing with the radii derived from the luminosities and photospheric temperature, 
we derive that $i$ is equal to 35\degr\ and 55\degr\ (to an accuracy of $\simeq$10\degr) for V819~Tau and V830~Tau respectively.  

Using the evolutionary models of \citet[][assuming solar metallicity and including convective overshooting]{Siess00}, 
we obtain that V819~Tau and V830~Tau are both $1.00\pm0.05$~\msun\ stars, and that their respective ages are $\simeq$3.2 and $\simeq$2.2~Myr 
(with relative error bars of $\simeq$50\%, see Fig~\ref{fig:hrd}), suggesting that V819~Tau may be a slightly evolved version (by about $\simeq$1~Myr) 
of V830~Tau\footnote{Using the most recent evolutionary models of \citet{Baraffe15}, we obtain slightly smaller masses (of 0.90~\msun) for both stars, along 
with smaller ages (of $\simeq$2.2 and $\simeq$1.4~Myr respectively).  We nevertheless kept estimates derived from \citet{Siess00} as our reference values for internal 
consistency with previous MaPP and MaTYSSE results.}.  This supports that disc dissipation for PMS stars in Taurus may occur on timescales as short as a few 
Myrs \citep{Williams11, Ingleby12}, even though 80--90\% of single stars in this star-formation region still host discs at similar ages \citep{Kraus12}.  
This makes V819~Tau and V830~Tau atypical in this respect, and thus of obvious interest for MaTYSSE.  
We also note that our two targets are both located close to the theoretical threshold at which 1~\msun\ stars cease to be fully convective (green line in Fig.~\ref{fig:hrd}).  
This may suggest that V830~Tau is still fully convective whereas V819~Tau has already developed a small radiative core;  our error bars on 
the location of both stars in the HR diagram are however still too large to reach a firm conclusion (see Fig.~\ref{fig:hrd}), to the point 
that even the opposite conclusion could actually be true.  Interestingly, V819~Tau and V830~Tau are also located close 
to the classical T~Tauri star GQ~Lup in the HR diagram \citep[][see circle in Fig.~\ref{fig:hrd}]{Donati12}, implying that their internal structures should be similar, 
if that of GQ~Lup is not significantly impacted by accretion \citep[only moderate for this cTTS, e.g.,][]{Donati12}.  
We summarize the main parameters of both stars in Table~\ref{tab:par}.  

Finally, we report that core emission is clearly detected in the \caii\ infrared triplet (IRT) lines of both V819~Tau and V830~Tau, with an average equivalent width of 
the emission core equal to $\simeq$15~\kms, corresponding to the amount expected from chromospheric emission for such PMS stars.  
Moreover, the \hei\ $D_3$ line is barely visible (average equivalent width of $\simeq$5~\kms) for both V819~Tau and V830~Tau, further demonstrating that 
accretion is no longer taking place on to their surfaces, in agreement with previous studies.  

\begin{figure*}
\hbox{\hspace{-4mm}
\includegraphics[scale=0.6,angle=-90]{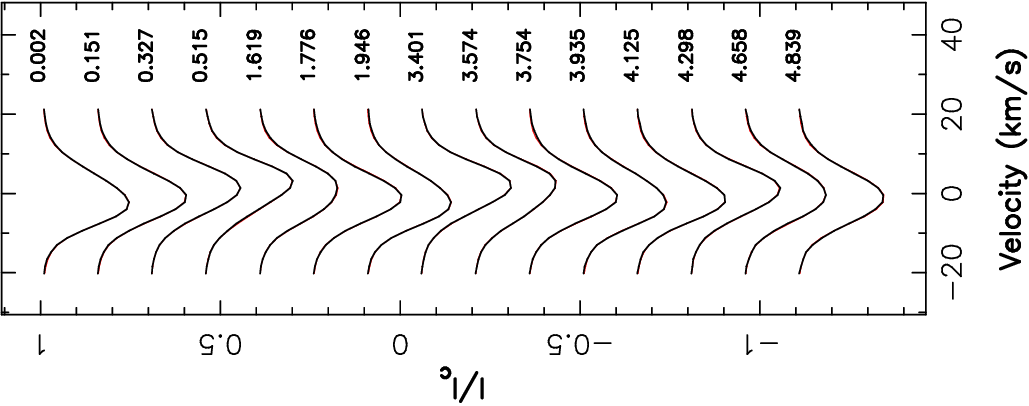}\hspace{3mm}
\includegraphics[scale=0.6,angle=-90]{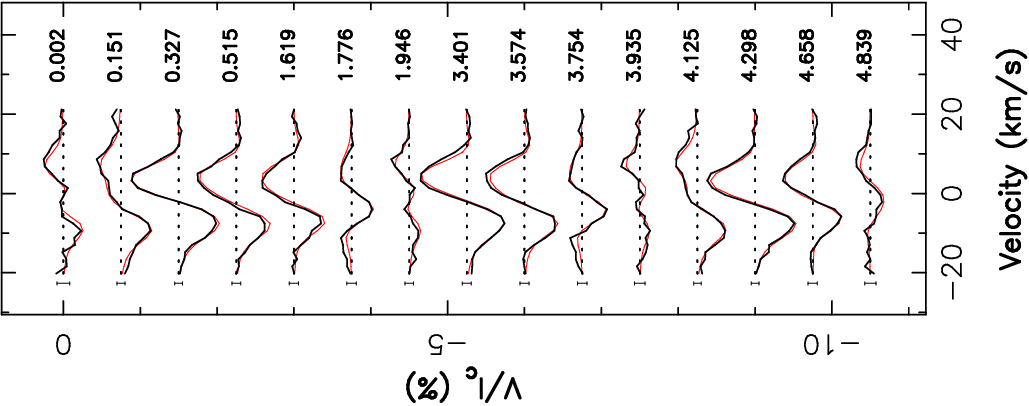}\hspace{6mm}
\includegraphics[scale=0.6,angle=-90]{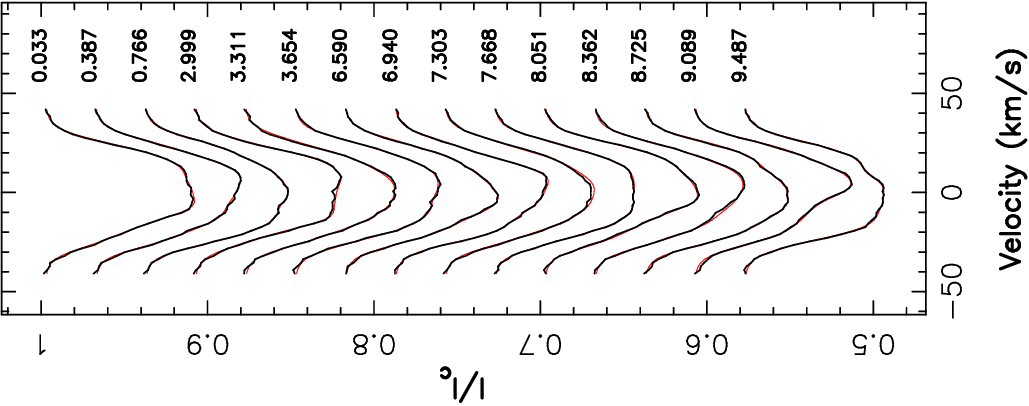}\hspace{3mm}
\includegraphics[scale=0.6,angle=-90]{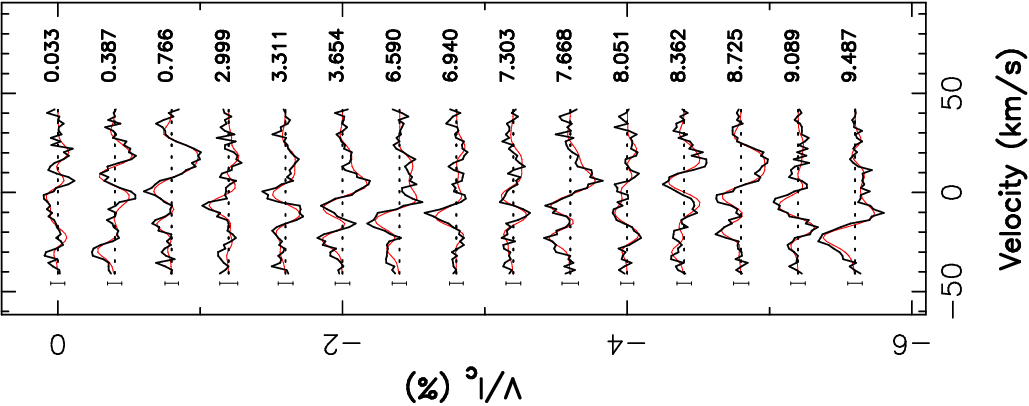}}
\caption[]{Maximum-entropy fit (thin red line) to the observed (thick black line) Stokes $I$ (first and third panels) 
and Stokes $V$ (second and fourth panels) LSD photospheric profiles of V819~Tau (first two panels) and V830~Tau (last two panels) in 2014~Dec and 2015~Jan.  
Rotational cycles and 3$\sigma$ error bars (for Stokes $V$ profiles) are also shown next to each profile. 
This figure is best viewed in color.  }
\label{fig:fit}
\end{figure*}

\section{Tomographic modelling}
\label{sec:mod}

Now that our two stars are well characterized regarding their atmospheric properties and evolutionary status, we are ready to apply our 
dedicated stellar-surface tomographic-imaging package to the spectropolarimetric data set described in Sec~\ref{sec:obs}.  
This tool is based on the principles of maximum-entropy image reconstruction and on the assumption that the observed variability is mainly 
caused by rotational modulation (with an added option for differential rotation);  this framework is well adapted to the case of wTTSs for 
which rotational modulation largely dominates intrinsic variability, in both photometry \citep[e.g.,][]{Grankin08} and photospheric lines 
\citep[e.g.,][]{Skelly10, Donati14}.  Since initially released \citep{Brown91, Donati97c}, the code underwent several upgrades \citep[e.g.,][]{Donati01, 
Donati06b}, the most recent ones being its re-profiling to the specific needs of MaPP and MaTYSSE observations \citep{Donati10b, Donati14}.  
More specifically, the imaging code is set up to invert (both automatically and simultaneously) time series of Stokes $I$ and $V$ LSD profiles 
into brightness and magnetic maps of the stellar surface;  moreover, brightness imaging is now allowed to reconstruct both cool spots and warm 
plages, known to contribute to the activity of very active stars \citep{Donati14}.  
The reader is referred to the papers mentioned above for more general details on the imaging method\footnote{As explained in \citet{Donati14}, 
full Stokes spectropolarimetry is not expected to bring much improvement in imaging quality over Stokes $I$ and $V$ spectropolarimetry, 
especially for stars as faint as our MaTYSSE targets.}.  

To compute the disc-integrated average photospheric LSD profiles, we start by synthesizing the local Stokes $I$ and $V$ profiles using Unno-Rachkovsky's 
analytical solution to the polarized radiative transfer equations in a Milne-Eddington model atmosphere, taking into account the local brightness and 
magnetic field;  we then integrate these local profiles over the visible hemisphere to retrieve the synthetic profiles to be compared with our observations.  
This computation method provides in particular a reliable description of how line profiles are distorted in the presence of magnetic fields \citep[including magneto-optical 
effects, e.g.,][]{Landi04}.  The main parameters of the local profile are similar to those used in our previous studies, the wavelength, Doppler width, equivalent 
width and Land\'e factor being respectively set to 670~nm, 1.8~\kms, 3.9~\kms\ and 1.2.  As part of the imaging process, we obtain accurate estimates for 
several parameters of both stars;  we find in particular that the \vrad's (the RVs the stars would have if unspotted) and \vsini's are respectively equal 
to $16.6\pm0.1$ and $9.5\pm0.5$~\kms\ for V819~Tau, and $17.5\pm0.1$ and $30.5\pm0.5$~\kms\ for V830~Tau (as listed in Table~\ref{tab:par}), in good 
agreement with previously published estimates \citep[e.g.,][]{Nguyen12}.  
We will revisit the RV curves of both stars in more details in Sec.~\ref{sec:fil}. 

\subsection{Brightness and magnetic imaging}

We show in Fig.~\ref{fig:fit} our sets of Stokes $I$ and $V$ LSD profiles of V819~Tau and V830~Tau along with our fits to the data.  
The fits we obtain correspond to a reduced chi-square \chisqr\ equal to 1 (i.e., a \chisq\ equal to the number of fitted data points, respectively equal 
to 360 and 705 for V819~Tau and V830~Tau), starting from initial values of about 65 and 10 (corresponding to null magnetic fields and unspotted brightness maps) 
for V819~Tau and V830~Tau respectively.  This further stresses the quality of our data set and the high performance 
of our imaging code at modelling the observed modulation of LSD profiles (also obvious from Fig.~\ref{fig:fit}).  The very good quality of the achieved fit 
clearly demonstrates that the assumption underlying our modelling, i.e., that the observed line profile variability is caused by the presence of 
brightness and magnetic surface structures rotating in and out of the observer's view, is verified.  As for LkCa~4 \citep{Donati14}, the reasonably dense phase 
coverage allows to track back most of the main reconstructed features into genuine profile distortions, and thus to safely claim that our maps include 
no major imaging artifact nor bias.  

\begin{figure*}
\hbox{\hspace{15mm}
\includegraphics[scale=0.45,angle=-90]{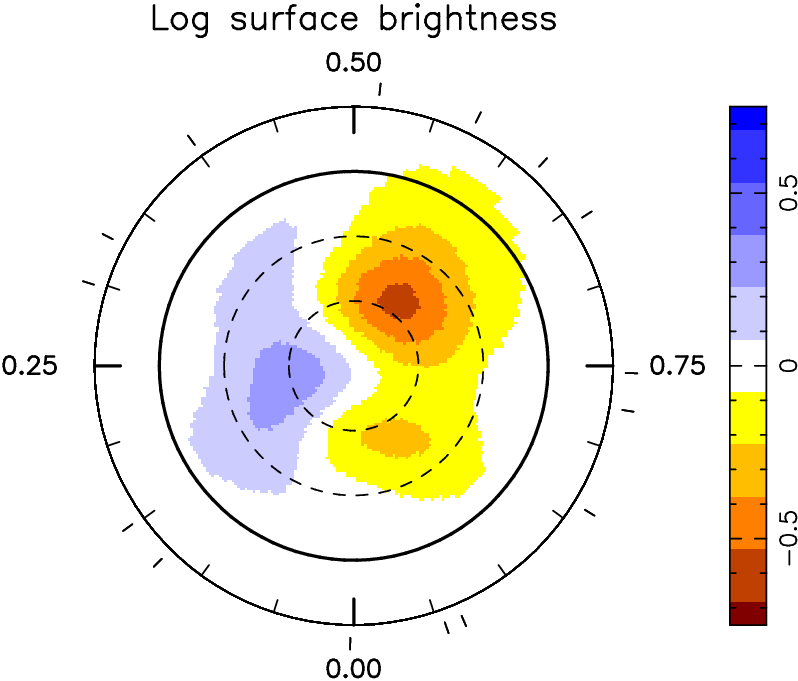}\hspace{20mm}
\includegraphics[scale=0.45,angle=-90]{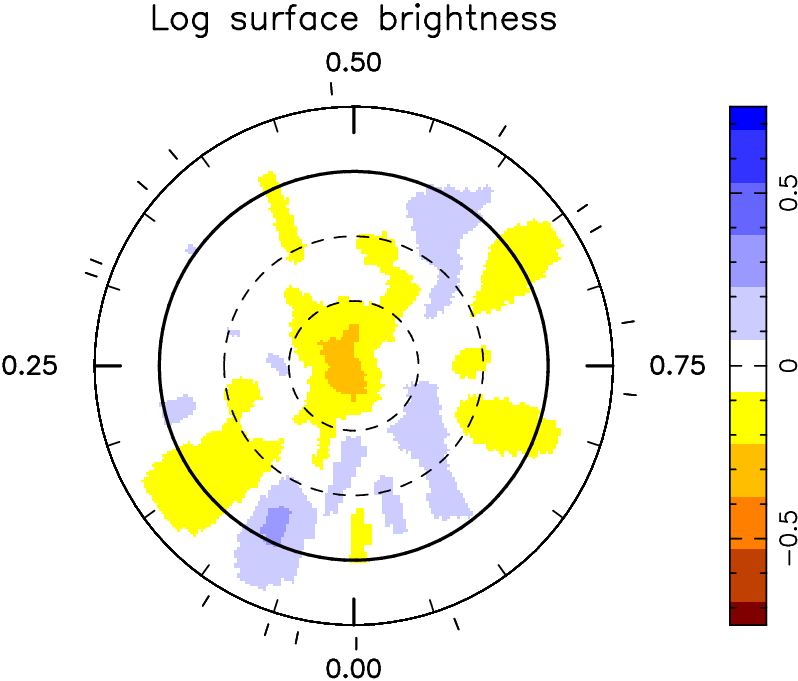}} 
\caption[]{Maps of the logarithmic brightness (relative to the quiet photosphere), 
at the surfaces of V819~Tau (left) and V830~Tau (right) in 2014~Dec -- 2015~Jan.    
The stars are  shown in flattened polar projection down to latitudes of $-30\degr$,
with the equator depicted as a bold circle and parallels as dashed circles.  Radial ticks around
each plot indicate phases of observations.  This figure is best viewed in color. }
\label{fig:mapi}
\end{figure*}

\begin{figure*}
\hbox{\includegraphics[scale=0.35,angle=-90]{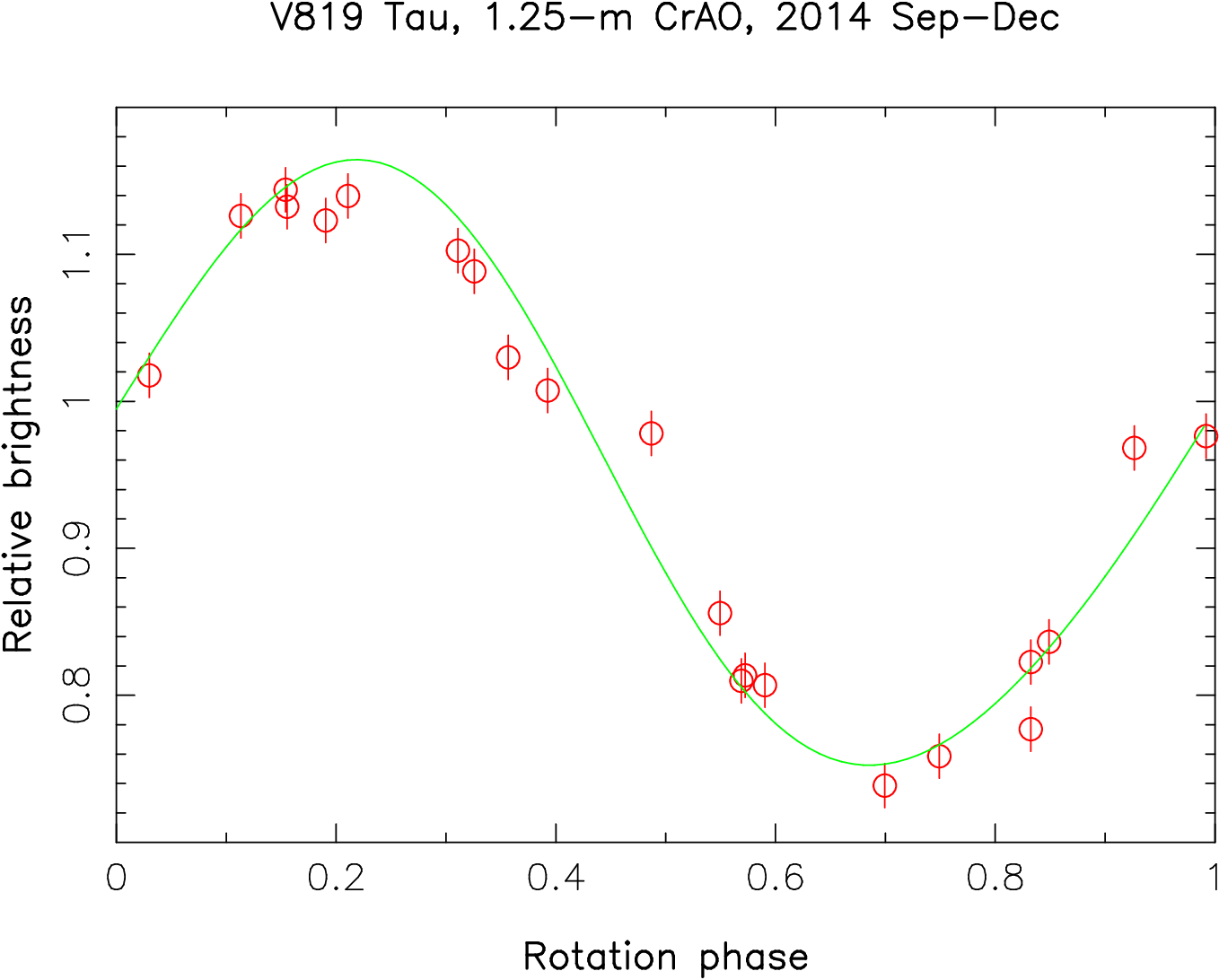}\hspace{6mm}\includegraphics[scale=0.35,angle=-90]{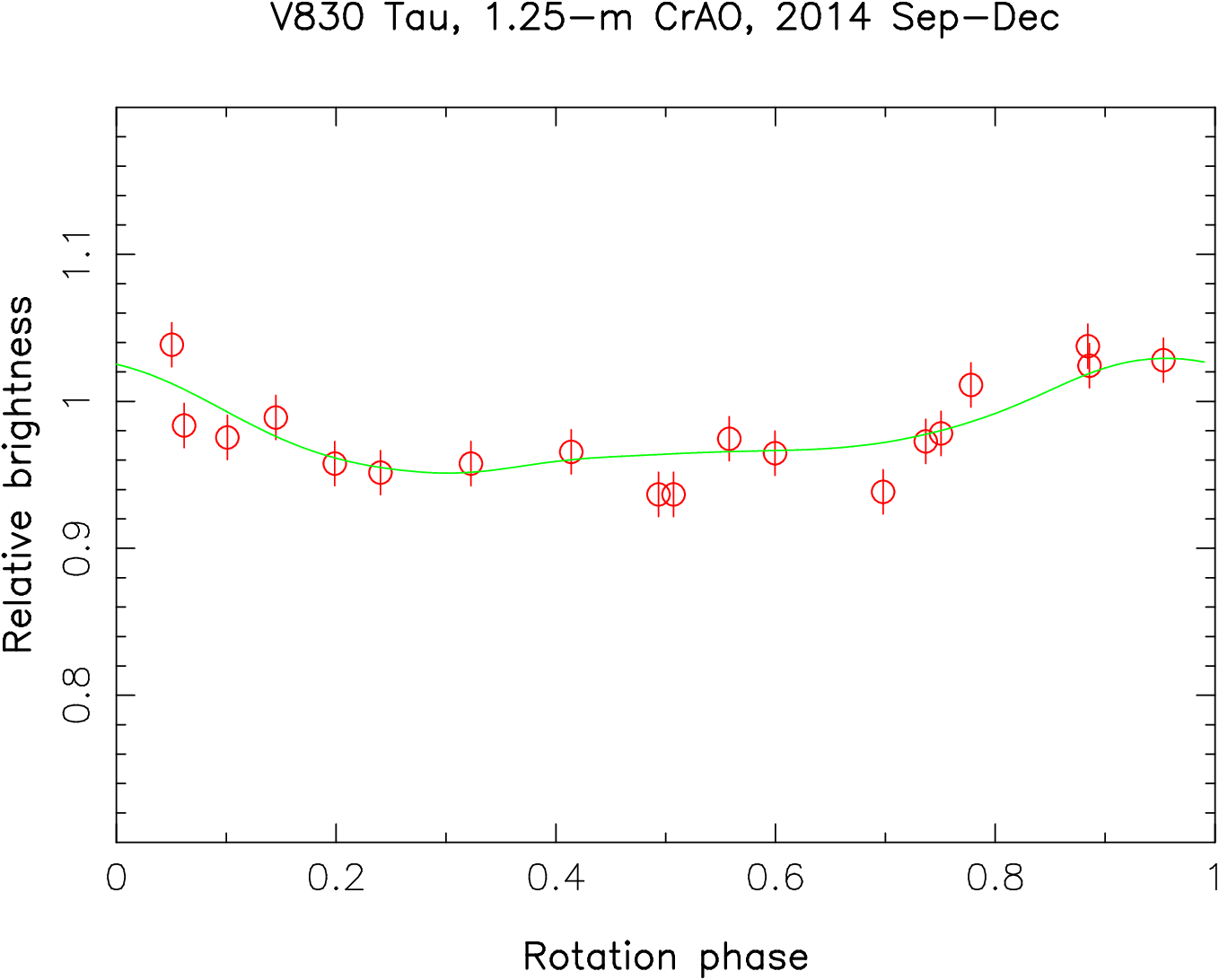}} 
\caption[]{Brightness variations of V819~Tau (left) and V830~Tau (right) predicted from the tomographic modelling of Fig.~\ref{fig:mapi}
of our spectropolarimetric data set (green line), compared with contemporaneous photometric observations in the \V\ band
(red open circles and error bars of 15~mmag) at the 1.25-m CrAO telescope (see Table~\ref{tab:pho}). }
\label{fig:pho}
\end{figure*}

The reconstructed brightness maps of V819~Tau and V830~Tau (see Fig.~\ref{fig:mapi}) both include cool spots and warm plages, but nevertheless 
feature a number of significant differences, regarding the contrast of these features in particular, much larger for V819~Tau than for V830~Tau.  
This difference is readily visible from our photometric data (collected at CrAO, see Table~\ref{tab:pho} and Fig.~\ref{fig:pho}) whose full amplitude 
is about 4.5$\times$ larger ($\simeq$0.45~mag) for V819~Tau than for V830~Tau ($\simeq$0.1~mag).  We emphasize that our reconstructed 
brightness images predict light curves that are in very good agreement with the observed ones (see Fig.~\ref{fig:pho}), even though these images were 
produced solely from our sets of LSD profiles;  this demonstrates that surface imaging from well-sampled series of high-quality spectroscopic data 
are capable of predicting photometric curves, at a typical relative precision of about 28 and 14~mmag for V819~Tau and V830~Tau respectively.  

The brightness map of V819~Tau (see Fig.~\ref{fig:mapi}, left panel) is reminiscent of that of LkCa~4 with a dark spot near the pole - even though this spot 
is not quite as dark and close to the pole as that of LkCa~4 \citep{Donati14};  it also comes with a warm plage next to it though this plage is again less bright 
than its equivalent on LkCa~4.  We suspect that part (but not all) of this difference reflects the lower \vsini\ of V819~Tau (with respect to LkCa~4) and thus the 
poorer spatial resolution achievable through Doppler imaging at the surface of this star.  As for LkCa~4, bright plages are again required to reproduce all details 
of the profile variability observed for V819~Tau, and in particular the large RV variations (of full amplitude $\simeq$2.4~\kms, see Sec.~\ref{sec:fil}) that V819~Tau 
exhibits.  The reconstructed brightness features are found to cover $\simeq$15\% of the overall stellar surface, with the main cool spot covering $\simeq$9\% 
and the plage $\simeq$6\%.  Note that these estimates of the spot / plage coverage should be considered as lower limits only (hence the larger spot coverage assumed 
in Sec.~\ref{sec:evo} to derive the location of both stars in the HR diagram);  Doppler imaging is indeed powerful at recovering large-scale features causing 
detectable profile distortions, but mostly insensitive to small-scale structures more or less evenly distributed over the stellar surface and generating no 
profile distortions.  

The brightness map of V830~Tau also features a cool spot close to the pole and at least one warm plage near the equator (see Fig.~\ref{fig:mapi}, right panel);  
although the overall surface covered with spots and plages on V830~Tau, equal to $\simeq$12\% ($\simeq$6\% for spots and $\simeq$6\% for plages), is comparable to 
that of V819~Tau, they feature a lower contrast in average and are spread over a more even range of longitudes, yielding a much flatter photometric response 
(see Fig.~\ref{fig:pho}, right panel) despite significant distortions of the LSD Stokes $I$ profiles (see Fig.~\ref{fig:fit}, third panel) and clear RV 
fluctuations (of full amplitude $\simeq$1.6~\kms, 
see Sec.~\ref{sec:fil}).  As already evidenced in the case of V410~Tau \citep[the first wTTS to be magnetically imaged, although with a slightly older version 
of our imaging code,][]{Skelly10}, we confirm that a flat photometric curve by no means implies the absence of spots (or the reduction in spot coverage) at the 
surface of the star, but rather a more even distribution of features (both cool and warm).  Moreover, the similar \vsini's of V830~Tau and LkCa~4 further argue 
(along with the different amplitudes of the photometric light curves) that the reconstructed brightness maps of these two stars trace genuine differences in the 
parent spot distributions at the surface of these two stars.  

\begin{figure*}
\vspace{15mm}
\hbox{\hspace{5mm}\includegraphics[scale=0.7,angle=-90]{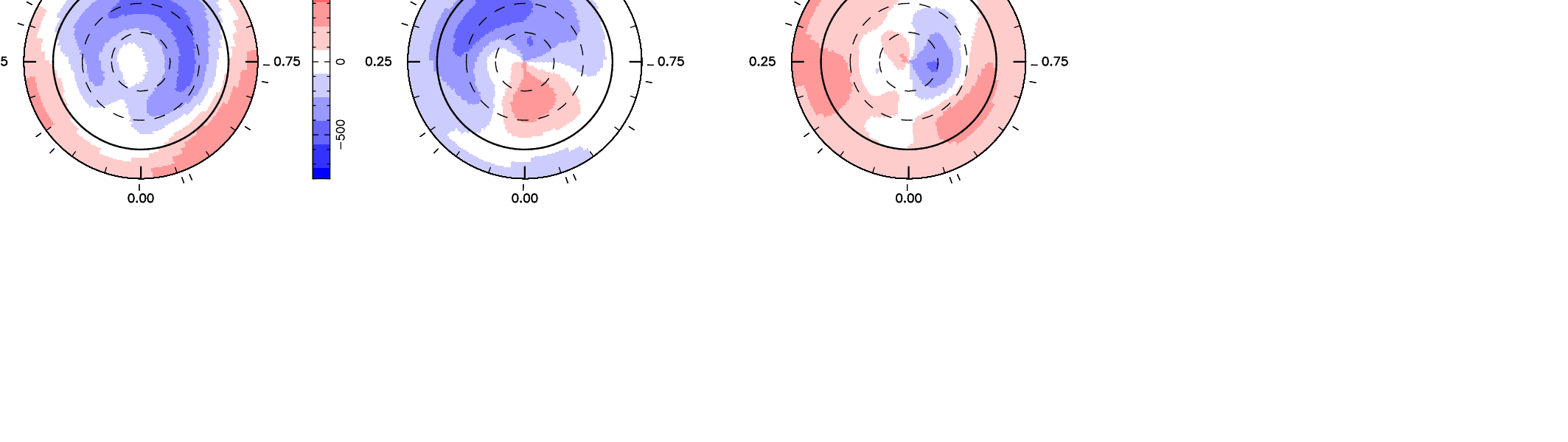}\vspace{-15mm}} 
\hbox{\hspace{5mm}\includegraphics[scale=0.7,angle=-90]{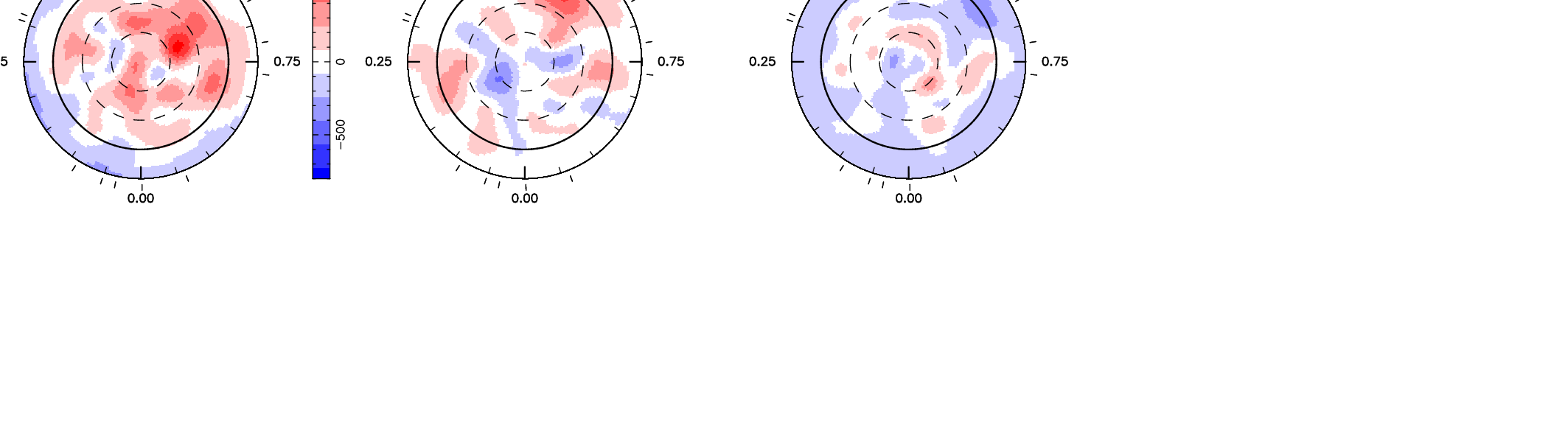}\vspace{-15mm}}
\vspace{-20mm}
\caption[]{Maps of the radial (left), azimuthal (middle) and meridional (right) components of the 
magnetic field $\bf B$ at the surfaces of V819~Tau (top) and V830~Tau (bottom) in 2014~Dec -- 2015~Jan.  
Magnetic fluxes in the color lookup table are expressed in G.  The stars are shown in flattened polar 
projection as in Fig~\ref{fig:mapi}.  This figure is best viewed in color. }
\label{fig:mapv}
\end{figure*}

\begin{figure*}
\hbox{\includegraphics[scale=0.52,angle=-90]{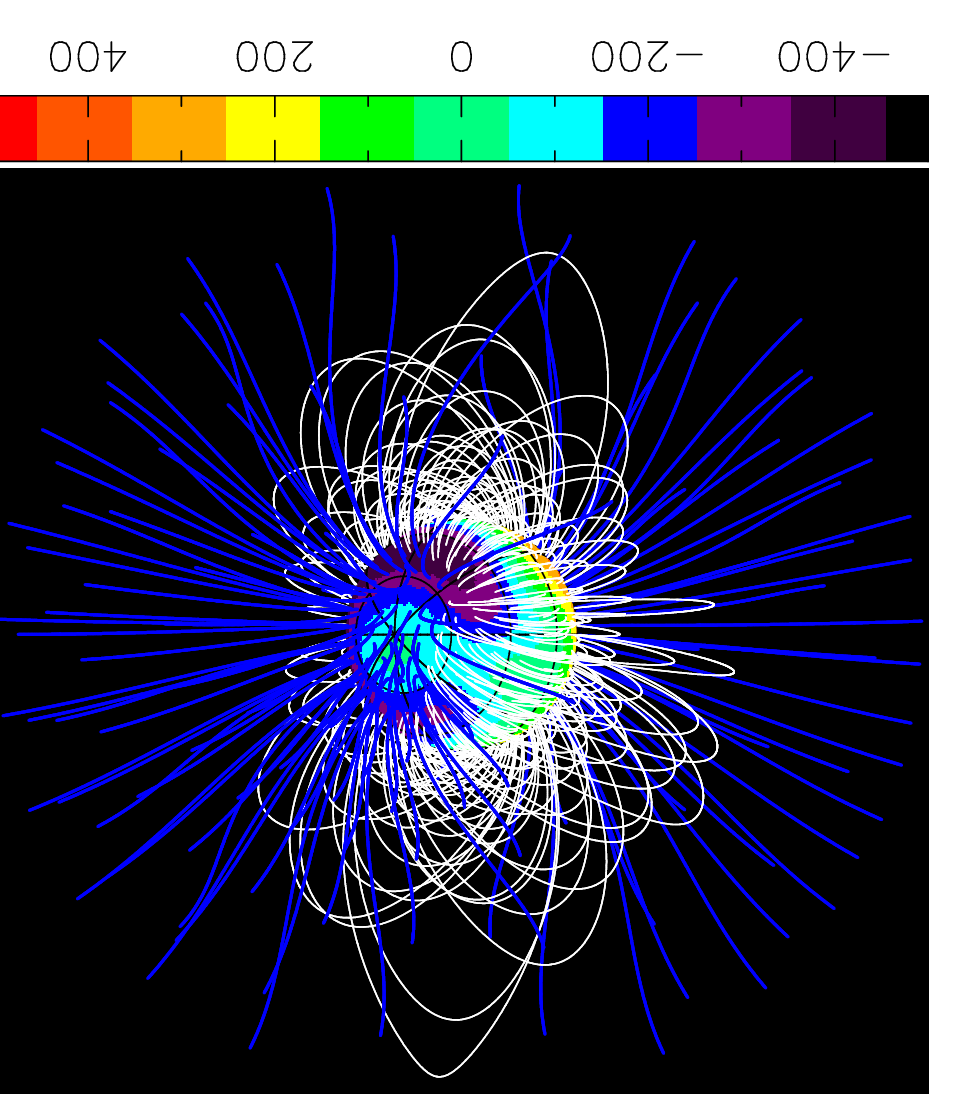}\includegraphics[scale=0.52,angle=-90]{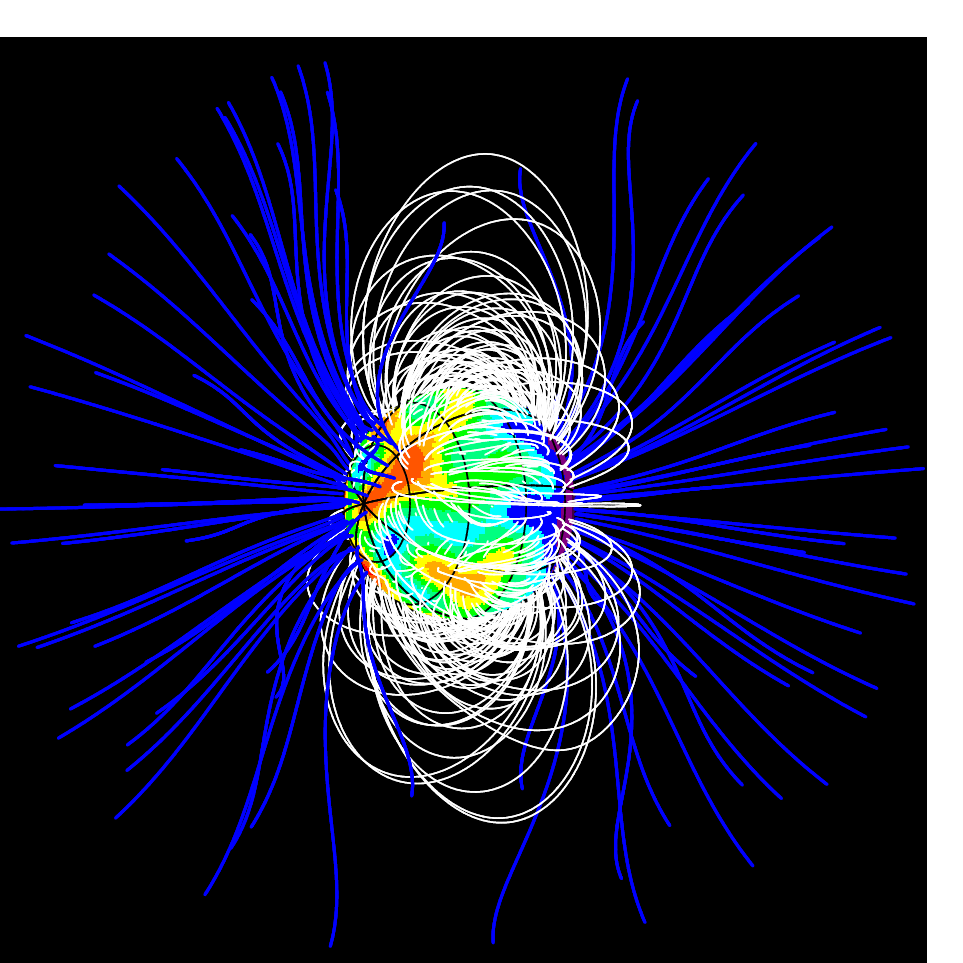}} 
\caption[]{Potential field extrapolations of the magnetic topologies reconstructed for V819~Tau (left) and V830~Tau (right), 
as seen by an Earth-based observer at phases 0.0 and 0.15 respectively.  Open and closed field lines are shown in blue and white respectively, 
whereas colors at the stellar surface depict the local values (in G) of the radial field fluxes (as shown in the left panels of Fig.~\ref{fig:mapv}).  
The source surface at which the field becomes radial is (arbitrarily) set at a (realistic) distance of 4~\rstar.  This figure is best viewed in color. }   
\label{fig:lsf}
\end{figure*}

Finally, we emphasize that the good agreement obtained for both stars between the predicted and observed light curves despite the large time span (of $\simeq$4~months) 
over which our photometric data were collected (as opposed to $\simeq$1~month only for our spectropolarimetric data, see Tables~\ref{tab:log} and \ref{tab:pho}) 
brings further evidence that surface features of wTTSs are stable and long-lived on a timescale of months, supporting previous conclusions from long-term photometric monitoring
\citep[e.g.,][]{Grankin08}.

The reconstructed magnetic fields of V819~Tau and V830~Tau (see Fig.~\ref{fig:mapv}) are found to be similar in terms of their intensities and topological 
properties, and are much weaker than the one found at the surface of LkCa~4 \citep[featuring kG poloidal and toroidal field components,][]{Donati14}.  
In both cases, the fields are mainly poloidal, at a level of 80\% and 90\% for V819~Tau and V830~Tau respectively, with average unsigned fluxes 
of 370 and 300~G.  Most of the reconstructed poloidal field energy (75\% and 60\%) concentrates in spherical harmonics (SH) dipolar modes (i.e., with
$\ell=1$, $\ell$ denoting the degree of the modes), whereas 90\% of it gathers in the aligned dipole mode ($\ell=1$ and $m=0$, with $m$ denoting the mode order) 
for both stars.  At first order and at some distance from the stars, the poloidal components of V819~Tau and V830~Tau can be 
approximated with dipoles of respective strengths 400 and 350~G, tilted at angles of $\simeq$30\degr\ to the line of sight towards phases 0.50 and 0.65\footnote{Note 
that these tilt and phase values refer to the visible magnetic pole, i.e., the negative and positive poles for V819~Tau and V830~Tau respectively.  Using the positive pole 
as reference \citep[as in][]{Gregory11}, the tilt angle and phase of the dipole component of V819~Tau become 150\degr\ and 0.0. }.  
A weaker octupolar component in the range 150--300~G \citep[depending on whether or not we favour odd modes in the imaging process, as for cTTSs,][]{Donati11, 
Donati12, Donati13}, is also present on V819~Tau, more or less aligned with the rotation axis and antiparallel with the dipole component;  
on V830~Tau, the octupolar component is weaker and non-axisymmetric, with a polar strength no larger than 150~G.  
The reconstructed large-scale fields of V819~Tau and V830~Tau also include weaker (though significant) toroidal components with average unsigned 
fluxes equal to $\simeq$170~G and $\simeq$100~G respectively, and whose topologies are more complex and less axisymmetric than that of LkCa~4 \citep{Donati14}.  

\begin{figure*}
\center{\hbox{\hspace{-2mm}
\includegraphics[scale=0.3,angle=-90]{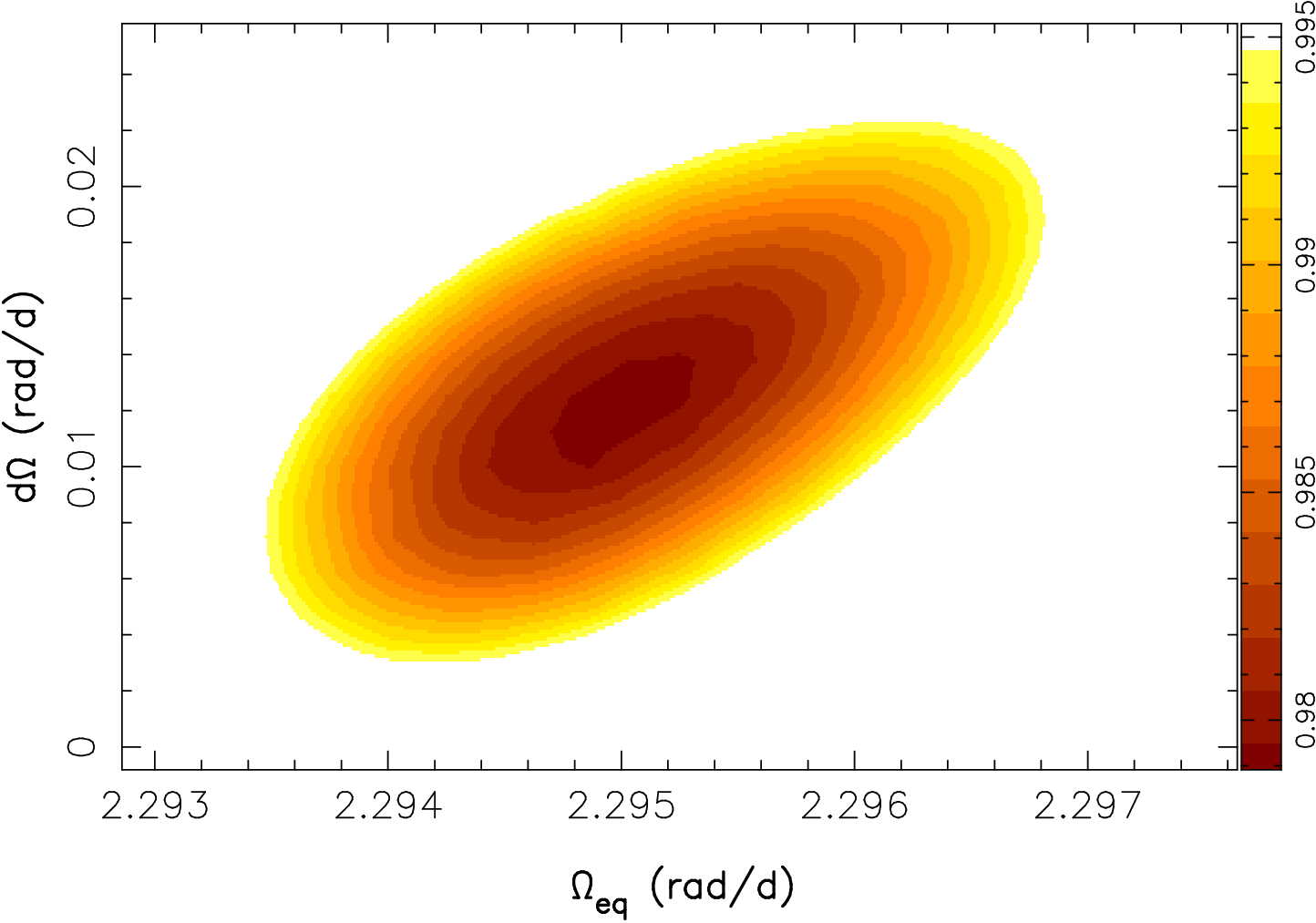}\hspace{2mm}
\includegraphics[scale=0.3,angle=-90]{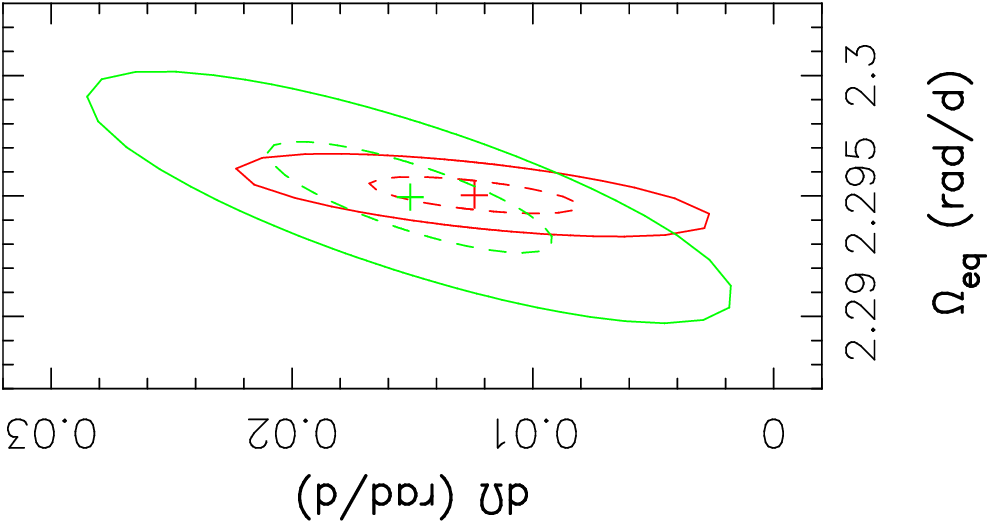}\hspace{2mm}
\includegraphics[scale=0.3,angle=-90]{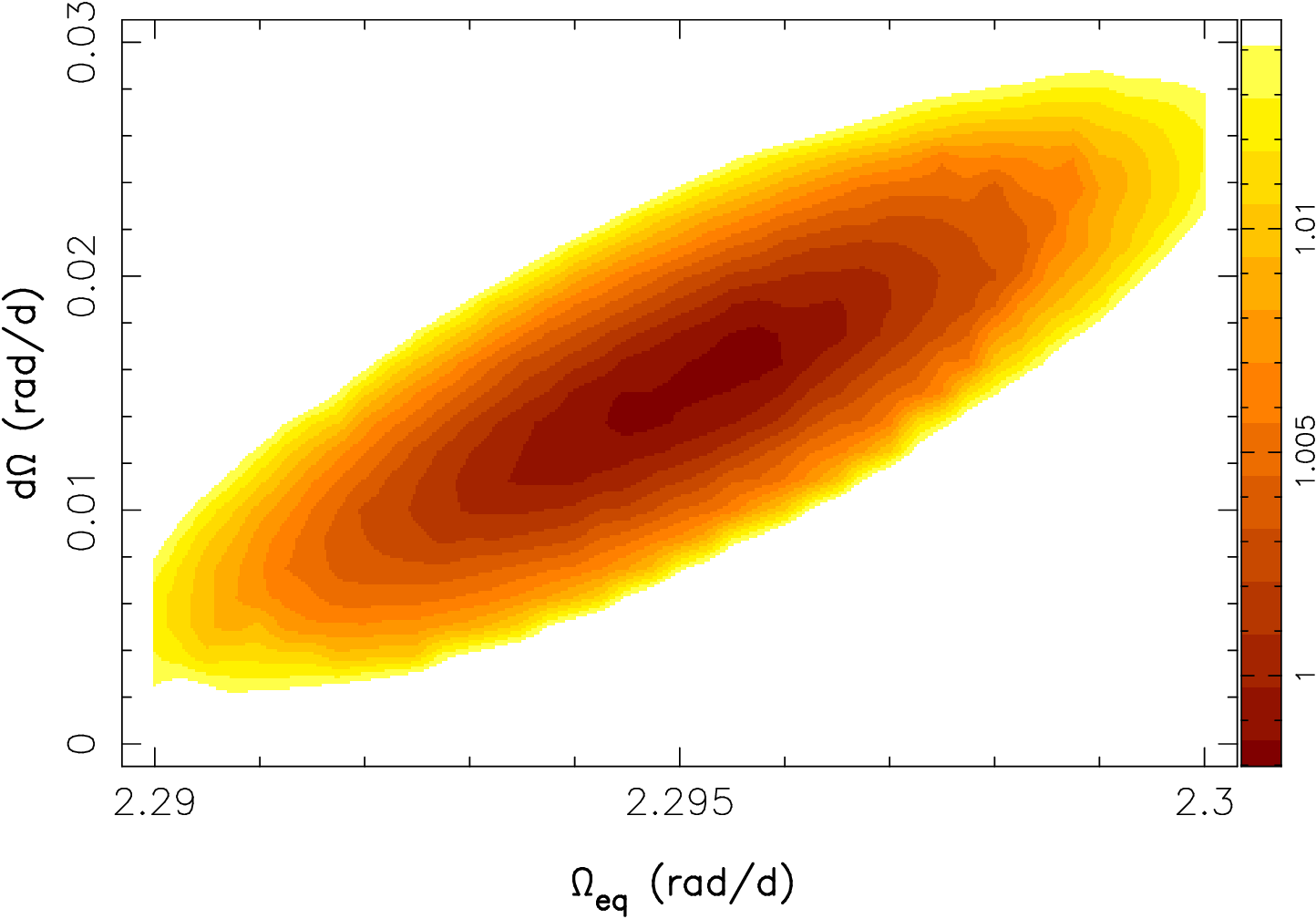}}} 
\caption[]{Variations of \chisqr\ as a function of \omeq\ and \dom, derived from the modelling of our Stokes $I$ (left) and $V$ (right) LSD profiles of V830~Tau at constant 
information content.  In both cases, a clear and well defined paraboloid is observed, with the outer color contour tracing the 1.7\% increase in \chisqr\ (or equivalently 
a \chisq\ increase of 11.8 for 705 fitted data points) that corresponds to a 3$\sigma$ ellipse for both parameters as a pair.  The middle plot emphasizes how well the 
confidence ellipses from both measurements overlap, with 1$\sigma$ and 3$\sigma$ ellipses (depicting respectively the 68.3\% and 99.73\% confidence levels) shown in full and 
dashed lines (in red and green for Stokes $I$ and $V$ data respectively).  This figure is best viewed in color.}
\label{fig:dr}
\end{figure*}

SH expansions describing the reconstructed field presented in Fig.~\ref{fig:mapv} are limited to terms with $\ell\leq8$ and 10 
for V819~Tau and V830~Tau respectively;  only marginal changes to the solution are observed when increasing the maximum $\ell$ value beyond these thresholds, 
demonstrating that most of the Zeeman signal detected in Stokes $V$ LSD profiles of both stars is captured in the images we recovered.  
As an example, we present in Fig.~\ref{fig:lsf} the extrapolated large-scale field topologies of V819~Tau and V830~Tau using the potential field approximation 
\citep[e.g.,][]{Jardine02a} and derived solely from the reconstructed radial field components;  as states of lowest possible magnetic energy, these potential 
topologies are shown to provide a reliable description of the magnetic field well within the Alfv\'en radius \citep{Jardine13}.  From these plots, the dominantly 
dipolar topologies of the large-scale fields of V819~Tau and V830~Tau are quite obvious.  

Finally, we note that the reconstructed brightness (see Fig.~\ref{fig:mapi}) and magnetic maps (see Fig.~\ref{fig:mapv}) 
of V819~Tau and V830~Tau only show a weak level of spatial correlation,  
as opposed to what we reported for LkCa~4 where the cool polar spot overlapped almost exactly with the region of strongest radial field \citep[see Fig.~4 of][]{Donati14}.  
The closest match of this kind that we can find in our new data is the cool spot close to the pole of V819~Tau, which roughly coincides with the visible pole of 
the dominant dipolar component of the large-scale field, but not with a local maximum of the surface field (as observed on LkCa~4).    
This is in line with what is observed on other cool active stars with large-scale field strengths similar 
to those of V819~Tau and V830~Tau \citep[e.g.,][]{Donati03a};  observations so far suggest that such tight spatial correlation between brightness and 
magnetic maps is only observed for young stars in which the large-scale poloidal field exceeds 1~kG \citep[like LkCa~4, but also, e.g., V410~Tau, GQ~Lup, 
V2129~Oph and DN~Tau,][]{Skelly10, Donati12, Donati11, Donati14}.

\subsection{Surface differential rotation}

Spread out over almost a full month (see Table~\ref{tab:log}), our spectropolarimetric data set allows us to 
estimate, at the surfaces of both stars, the amount of latitudinal differential rotation shearing the brightness and / or magnetic maps.  
We achieve this in practice by assuming that the rotation rate at the surface of the star is varying with latitude $\theta$ as  $\omeq - \dom \sin^2 \theta$
where \omeq\ is the rotation rate at the equator and \dom\ the difference in rotation rate between the equator and the pole;  in this context, 
one can measure differential rotation by finding out the pair of parameters \omeq\ and \dom\ that produces the brightness / magnetic map with minimum information content.  
This procedure was successfully used in a large number of studies over the last 15 yrs \citep[e.g.,][]{Donati00, Petit02, Donati03b}, 
including on two wTTSs \citep{Skelly10, Donati14}.  The reader is referred to these papers for more information on this technique.  

Thanks to its high \vsini\ (see Table~\ref{tab:par}) and the good spatial resolution it offers, V830~Tau is an ideal target for estimating differential rotation;  
moreover, the presence 
of surface features (in both the brightness and magnetic maps) at various latitudes (see Figs.~\ref{fig:mapi} and \ref{fig:mapv}) provides adequate information 
to obtain an accurate measurement.  The \chisqr\ maps we obtain (as a function of both \omeq\ and \dom) are shown in Fig.~\ref{fig:dr};  
the one derived from our Stokes $I$ data features a clear minimum at $\omeq=2.2950\pm0.0005$~\rpd\ and $\dom=0.0124\pm0.0029$~\rpd\ (left panel of 
Fig.~\ref{fig:dr}), whereas that inferred from our Stokes $V$ data fully supports this estimate though with larger error bars 
($\omeq=2.2950\pm0.0015$~\rpd\ and $\dom=0.0152\pm0.0039$~\rpd, right panel of Fig.~\ref{fig:dr}).  
These values translate into rotation periods at the equator and pole of $2.738\pm0.001$~d and $2.753\pm0.004$~d respectively, fully consistent
with the time-dependent periods of photometric fluctuations reported in the literature for V830~Tau \citep[ranging from 2.738 to 2.747~d when considering only 
definite detections of photometric modulation,][]{Grankin08, Xiao12} known to also probe surface differential rotation. 

For V819~Tau, the spatial resolution of our maps is significantly coarser due to the lower \vsini\ (see Table~\ref{tab:par});  
as a result, our Stokes $I$ data yield a \chisqr\ map 
showing no clear minimum, preventing us from estimating differential rotation.  This is actually not very surprising given that the brightness map we 
recovered only includes two main features (one cool spot and one warm plage) more or less located at the same latitude (see Fig.~\ref{fig:mapi});  our data is 
thus lacking information about rotation periods and recurrence rates of profile distortions at different latitudes, without which differential rotation 
cannot be estimated \citep{Petit02}.  Our Stokes $V$ data set yield a more favourable situation and allow us to obtain an estimate 
of surface differential rotation, equal to  $\omeq=1.141\pm0.004$~\rpd\ and $\dom=0.008\pm0.008$~\rpd;  unsurprisingly, the precision of this estimate is 
worse than for that of V830~Tau, reflecting the coarser spatial resolution achievable for this more slowly rotating star.  
These values translate into rotation periods at the equator and pole of $5.51\pm0.02$~d and $5.55\pm0.04$~d respectively.  Our estimate is reasonably consistent with 
most photometric periods reported in the literature for V819~Tau \citep[ranging from 5.50 to 5.61~d,][]{Grankin08, Xiao12} except for one discrepant value 
\citep[of 5.73~d, derived from data collected in 2004,][]{Grankin08}.  Looking back at the original photometric data, we find that this discrepant value 
actually corresponds to a low-accuracy estimate, reflecting the small number of measurements and the short observation time span at this specific epoch.  

Our results indicate that surface differential rotation is small on both V819~Tau and V830~Tau, much weaker in particular than that of the Sun (equal to $\simeq0.055$~\rpd).  
Whereas we cannot exclude (from our data alone) that V819~Tau rotates as a solid body, the measurement we obtain for V830~Tau indicates that it truly rotates differentially 
and that the amount of surface shear that it suffers is $\simeq$4.4$\times$ weaker than that of the Sun, with a typical time of $\simeq$500~d for the 
equator to lap the pole by one complete rotation cycle.  Similarly small amounts of surface differential rotation were reported for the fully-convective wTTSs LkCa~4 and 
V410~Tau \citep{Donati14, Skelly10} as well as for fully convective M dwarfs with presumably similar internal structures \citep{Donati06a, Morin08a}.  

\begin{figure*}
\hbox{\includegraphics[scale=0.35,angle=-90]{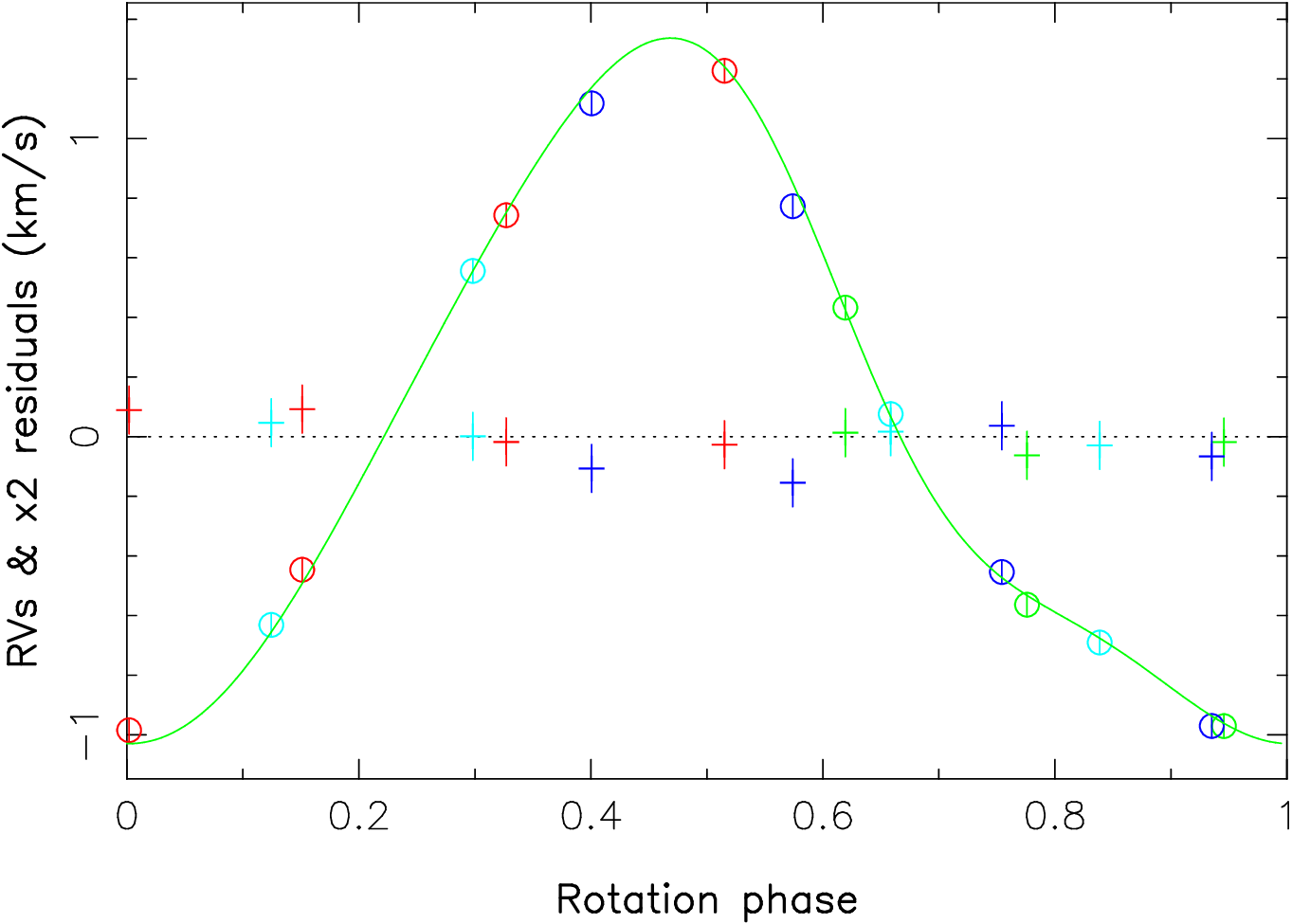}\hspace{6mm}\includegraphics[scale=0.35,angle=-90]{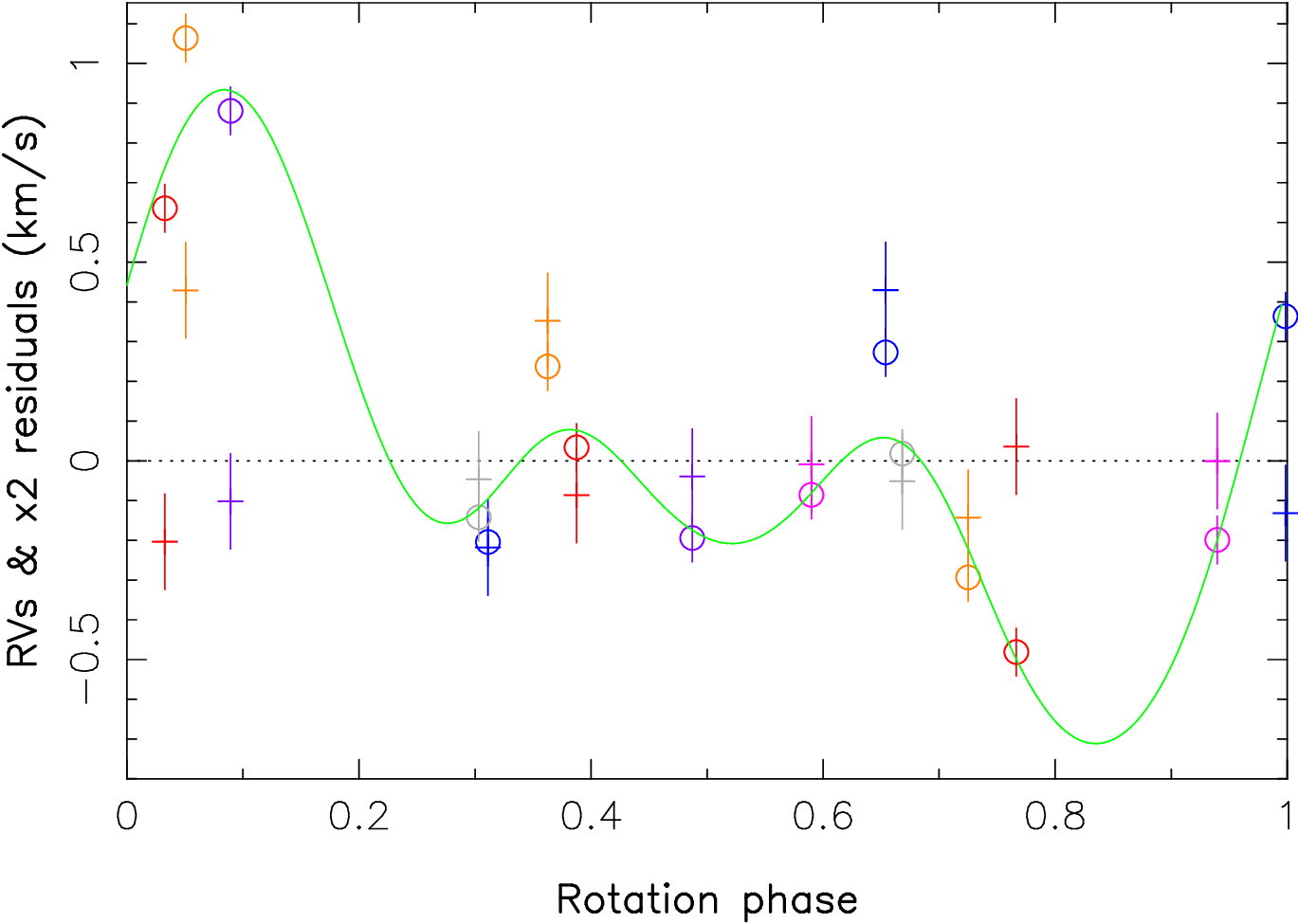}} 
\caption[]{RV variations (in the stellar rest frame) of V819~Tau (left) and V830~Tau (right) as a function of rotation phase, as measured from our observations 
(open circles) and predicted by the tomographic maps of Fig.~\ref{fig:mapi} (green line).  RV residuals (expanded by a factor of 2 for 
clarity), are also shown (pluses) and exhibit a rms dispersion equal to 0.033~\kms\ for V819~Tau and 0.104~\kms\ for V830~Tau.  
Red, green, dark-blue, light-blue, magenta, grey, orange and purple symbols depict measurements secured at rotation cycles 0, 1, 3 (including phase 2.999), 4, 6, 7, 8 
and 9 respectively (see Table~\ref{tab:log} and Fig.~\ref{fig:fit}).  Given the asymmetric and often irregular shapes of Stokes $I$ LSD profiles 
(see Fig.~\ref{fig:fit}), 
RVs are estimated as first order moments of the Stokes $I$ LSD profiles rather than through Gaussian fits.  RV estimates and residuals are depicted with error bars of 
$\pm$0.04 and $\pm$0.06~\kms\ for V819~Tau and V830~Tau respectively.  This figure is best viewed in color.}
\label{fig:rv}
\end{figure*}

\begin{figure}
\includegraphics[scale=0.35,angle=-90]{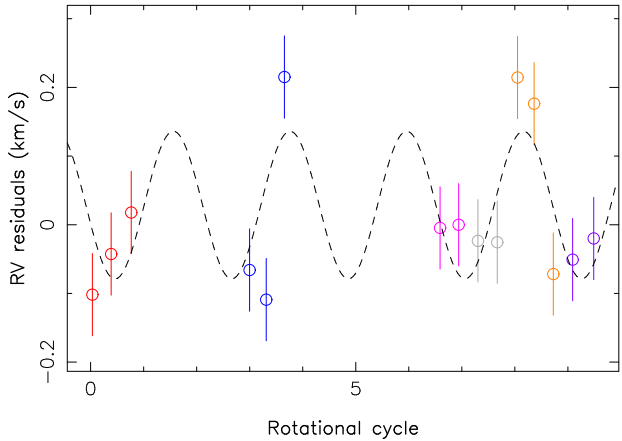} 
\caption[]{RV residuals of V830~Tau after applying our filtering process to the spectropolarimetric data set.  A sine fit to the points with an orbital period 
of $2.2\Prot\simeq6.0$~d and a velocity semi-amplitude of 0.11~\kms\ (dashed line) provides a good match to the observations 
(rms dispersion of 0.074~\kms), suggesting the possible presence of a giant planet orbiting close to V830~Tau.  More observations are obviously 
needed to confirm or reject this preliminary result.  The color code used for this plot is the same as that of Fig.~\ref{fig:rv} (right panel).  
This figure is best viewed in color.  } 
\label{fig:res}
\end{figure}

\section{Filtering the activity jitter}
\label{sec:fil}

As previously achieved for the wTTS LkCa~4 \citep{Donati14}, we propose to exploit the results of surface imaging to filter out the activity jitter in the RV curves of 
V819~Tau and V830~Tau, and potentially reveal the presence of close-in giant planets \citep[see][for more details]{Donati14}.
Technically speaking, the goal is to use the brightness and magnetic maps reconstructed from our 
phase-resolved spectropolarimetric data sets with tomographic imaging, to obtain an accurate description of the activity jitter that plagues the RV curves of 
our wTTSs;  once RV curves are filtered out from the predicted activity jitter, one can look for periodic signals in the RV residuals that may probe the presence of 
hJs.  Alternatively, we can {\em simultaneously} look for the presence of a planet while carrying out 
the imaging process, as recently proposed by \citet{Petit15};  in this case, the goal is to find out the optimal planet parameters, defined as those 
yielding the minimum amount of surface features for a given \chisqr\ fit to the data.  Our initial simulations suggest that both techniques should allow one to detect 
potential hJs orbiting wTTSs (provided their periods are not too close to the stellar rotation period or its first harmonics) and to quantitatively investigate 
whether close-in giant planets are much more frequent around low-mass forming stars at the TTS stage than around mature MS stars.  
Ultimately, our goal is to assess the likelihood of the core accretion coupled to disc 
migration scenario in which giant protoplanets mostly form in outer accretion discs then migrate inwards until they stop at a distance of a few hundredth of an AU, e.g., 
once they enter the central magnetospheric gaps of cTTSs and no longer experience the inward torque from the disc. 

The predicted activity jitter and filtered RV curves we derive for V819~Tau and V830~Tau are shown in Fig.~\ref{fig:rv}.  
Note that for our study, RVs are estimated as first order moments of the Stokes $I$ LSD profiles rather than through Gaussian fits, given the asymmetric and often 
irregular shapes of spectral lines (see Fig.~\ref{fig:fit}, first and third panels).  The filtering process is found to be quite efficient 
for V819~Tau, with RV residuals exhibiting a rms dispersion of only 0.033~\kms\ (for a full RV amplitude prior to the filtering process equal to 
$\simeq$1.6~\kms).  The error bar on these measurements is conservatively set to $\pm$0.04~\kms, reflecting mostly the intrinsic RV precision 
of ESPaDOnS \citep[of 0.03~\kms\ 
rms, e.g.,][]{Moutou07, Donati08b} and in a smaller part the intrinsic error of our filtering process (scaling up with the width of spectral lines, and estimated to be 
0.02~\kms\ rms in the present case).  This suggests that V819~Tau is unlikely to host a hJ with an orbital period in the range of 
what we can detect \citep[i.e.\ not too close 
to the stellar rotation period or its first harmonics, see][]{Donati14}, with a 3$\sigma$ error bar on the semi-amplitude of the RV residuals equal to 
0.07~\kms.  

For V830~Tau however, the RV curve exhibits significant scatter at several rotation phases (0.05, 0.35 and 0.65, see Fig~\ref{fig:rv} right panel), causing the 
RV residuals after filtering to reach a rms dispersion of 0.104~\kms\ (i.e., about twice and thrice those reported for LkCa~4 and V819~Tau respectively) and to exceed the 
value (of $\simeq$0.1~\kms) that potentially suggests the presence of a hJ \citep[according to the results of our preliminary simulations, see][]{Donati14}.  
The error bar on our individual RV residuals is set to $\pm$0.06~\kms, identical to those of LkCa~4 given the very similar widths of spectral lines and noise levels in our 
data.   A closer look reveals that this enhanced dispersion is mainly caused by 3 points (at rotational cycles 3.654, 8.051 and 8.362, see Fig.~\ref{fig:res}) that 
depart (RV-wise) from the bulk of our observations.  The temporal variations of the RV residuals are compatible with a sine wave with a period of either $4.0\pm0.2$~d or 
$6.0\pm0.2$~d and a semi-amplitude of $0.11\pm0.03$~\kms.  Bringing a \chisq\ improvement of $\simeq$22 for 4 degrees of freedom, this fit suggests that the RV residuals do 
indeed contain a RV signal, with a confidence level of 99.98\%.  However, given the very limited sampling, we cannot firmly conclude that the RV signal we find 
is truly periodic as expected for a signal of planetary origin, the false alarm probability on the period detection 
(as derived from a Lomb-Scargle periodogram) being uncomfortably high (at about 35\%).  

\begin{figure}
\includegraphics[scale=0.35,angle=-90]{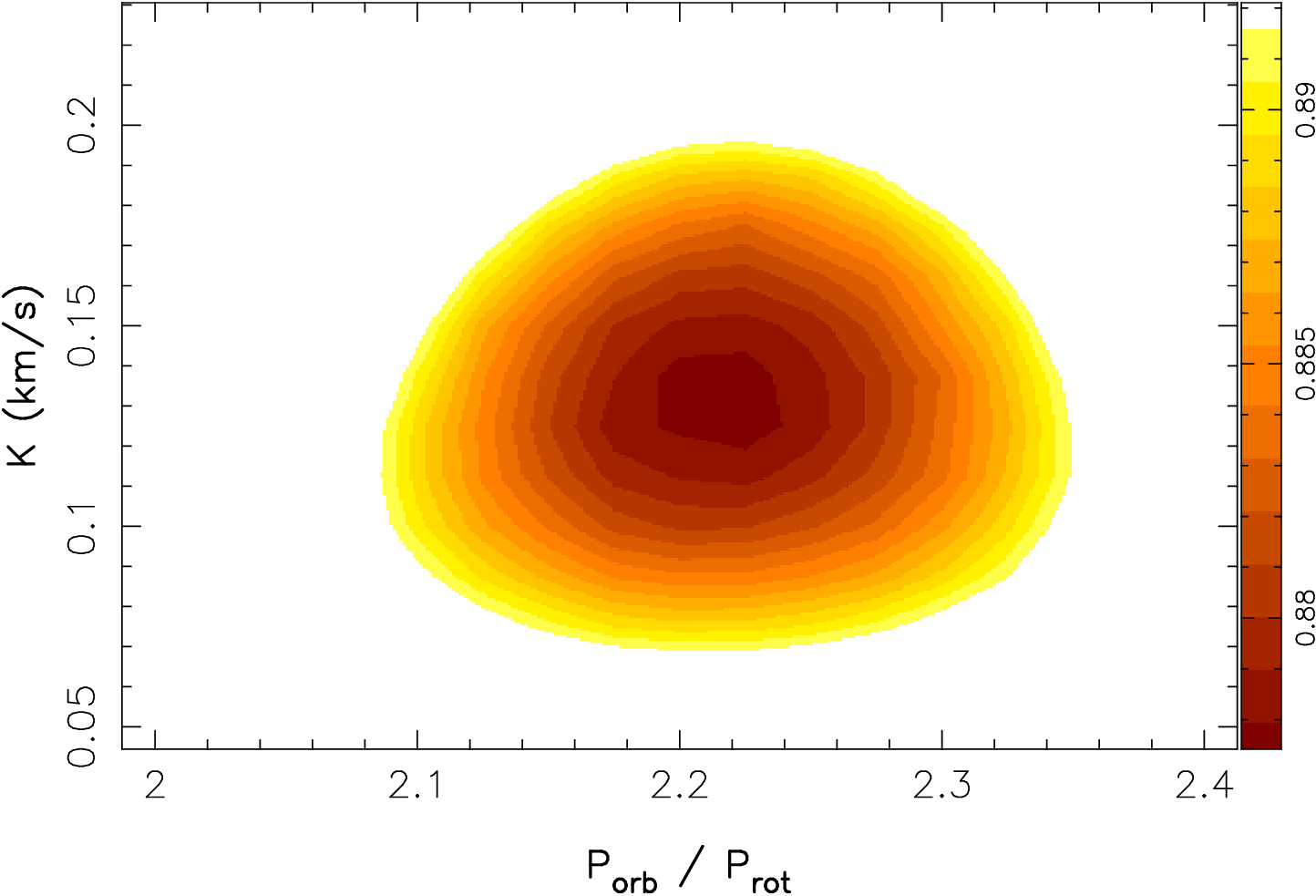} 
\caption[]{Variations of \chisqr\ as a function of $\Porb/\Prot$ and $K$, derived from the simultaneous modelling of the brightness features at the surface of V830~Tau and 
the orbital parameters of a close-in giant planet in circular orbit, following the new technique proposed by \citet{Petit15}.  A clear minimum is obtained in the \chisq\ 
landscape, whose projection in a $K$ vs $\Porb/\Prot$ plane passing through the minimum is shown here, with the outer color contour delimiting the 3$\sigma$ (or 99.73\%) 
confidence region.  This result suggests that a $\simeq$1.4~\mjup\ planet may be orbiting around V830~Tau.  This figure is best viewed in color. }   
\label{fig:pla}
\end{figure}

Experimenting the new technique proposed by \citet{Petit15}, whose advantage is to model both surface features and orbital parameters simultaneously, and thus to minimize 
crosstalk between the two and to maximize the precision of the whole process (at the expense of more computing time), we further confirm our finding, and conclude that the 
best tentative period of the signal in the RV residuals is $\Porb=6.0\pm0.2$~d, whereas the other option ($\Porb=4.0\pm0.2$~d) yields a worse fit to the data and is 
therefore less likely.  The semi-amplitude of the residual RV signal that we derive in this case (taking into account differential rotation as measured 
in Sec.~\ref{sec:mod}) is equal to $K=0.13\pm0.02$~\kms, consistent with, and presumably more accurate than, our initial estimate (of $0.11\pm0.03$~\kms).  
A projection of the achieved \chisqr\ landscape in the $K$ vs $\Porb/\Prot$ plane is shown in Fig.~\ref{fig:pla}, demonstrating that our solution corresponds to a well 
defined minimum.  The associated improvement in \chisq\ (with respect to a solution with no RV signal, see Fig.~\ref{fig:dr}, left panel) is larger than 70 
(for 4 degrees of freedom), implying a definite detection of a RV signal, with a very small false alarm probability ($<10^{-10}$)\footnote{This false alarm probability 
is smaller than that obtained by directly fitting RVs (equal to 0.02\%, see previous paragraph), which confirms that line profiles contain more information than 
their first moments.  Note that both tests only refer to the detection of an RV signal and not to its periodic nature (for which the false alarm probability is much larger).  }.

This residual RV signal, if confirmed, would be attributable to a $\simeq$1.4~\mjup\ planet located at a distance of $\simeq$0.065~AU from V830~Tau, 
assuming a circular orbit with an orbital plane coinciding with the equatorial plane of V830~Tau.  As expected, the impact of this residual RV signal on 
the reconstructed maps of Figs.~\ref{fig:mapi} and \ref{fig:mapv} is small, implying in particular that all conclusions of Sec.~\ref{sec:mod} are unaffected;  
this also applies to our differential rotation estimates of V830~Tau, that are not modified by more than a fraction of an error bar once the residual RV signal 
is removed, even for our most accurate measurement from Stokes $I$ data.
Similarly, the 3$\sigma$ upper limit on the semi-amplitude of the RV residuals of V819~Tau (equal to 0.07~\kms) translates into a value of $\simeq$1.3~\mjup\ 
for the mass of a planet orbiting at a distance of $\simeq$0.1~AU (and assuming again a circular orbit in the equatorial plane of the star).  

\begin{figure}
\includegraphics[scale=0.35,angle=-90]{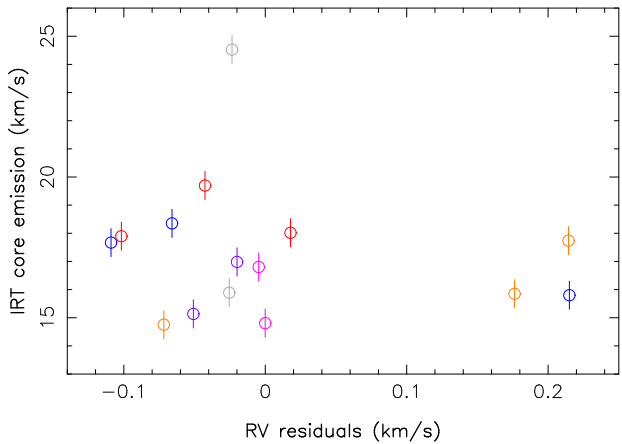} 
\caption[]{Equivalent width of the emission component in the core of \caii\ IRT lines (in \kms) as a function of RV residuals for V830~Tau.  No clear correlation is observed 
between both quantities, indicating that the observed RV residuals are likely not attributable to non-rotationally-modulated activity.  Note in particular that the 3 
highest RV residuals all correspond to average levels of IRT core emission, whereas the highest level of IRT core emission (presumably probing a small flaring event) 
corresponds to an average RV.  The color code used for this plot is the same as that of Fig.~\ref{fig:rv} (right panel) and Fig.~\ref{fig:res}.  This figure is best viewed in color.   }  
\label{fig:irt}
\end{figure}

We further checked that the temporal variations of the RV residuals we report here do not correlate with activity (e.g., estimated from the core emission in 
\caii\ IRT lines, see Fig.~\ref{fig:irt}), 
indicating that it cannot be attributed to, e.g., flaring events that may have damaged the profile shapes and corresponding RVs in our spectra.  Similarly, 
we emphasize that these RV variations do occur on relatively short timescales of only $\simeq$1~d, (e.g., from rotational cycle 3.311 to 3.654, and 8.362 to 8.725), 
suggesting that they are not attributable to some (slower) evolution of the brightness distribution at the surface of V830~Tau.  
Finally, we outline that RV variations are observed for V830~Tau 
only and not for V819~Tau, suggesting that they are likely not attributable to instrumental effects, that would equally plague spectra of both stars.  
We nevertheless stress that the detection of this RV signal currently relies on a small number of points, preventing us in particular to firmly assess its 
periodic nature;  as a result, we cannot yet entirely exclude the option that it reflects some yet unclear systematics in our data.  

\section{Summary \& discussion}
\label{sec:dis}

In this paper, we report results from a new set of spectropolarimetric observations collected with ESPaDOnS at the CFHT on two wTTSs, V819~Tau and V830~Tau, 
from mid 2014 December to mid 2015 January, and complemented by contemporaneous photometric observations from the \hbox{1.25-m} telescope at CrAO.  As for our 
initial study on LkCa~4 \citep{Donati14}, the monitoring of these two wTTSs was carried out in the framework of the international MaTYSSE Large Programme.  

From the analysis of our spectra, we find that both 
stars have similar atmospheric properties (photospheric temperature of $4250\pm50$~K and logarithmic gravity in cgs units of $3.9\pm0.1$), suggesting that 
V819~Tau and V830~Tau are $1.00\pm0.05$~\msun\ stars with respective radii of $1.8\pm0.2$~\rsun\ and $2.0\pm0.2$~\rsun\ viewed at inclination angles 
of 35\degr\ and 55\degr.  With estimated ages of $\simeq$3.2 and $\simeq$2.2~Myr, V819~Tau and V830~Tau are both close to the boundary 
at which 1~\msun\ stars start to build up a radiative core - with V819~Tau likely showing up as a slightly more evolved version than V830~Tau (note however 
the significant uncertainties on the ages, see Fig.~\ref{fig:hrd}).  V819~Tau and V830~Tau also lie close to GQ~Lup in the HR diagram \citep{Donati12}, although 
they belong to a different star-formation region.  

With rotation periods of 5.53~d and 2.74~d, V819~Tau and V830~Tau both spin significantly faster than typical cTTSs of similar mass (rotating in $\simeq$8~d, 
like GQ~Lup);  this suggests that both stars already entered a process of rapid spin-up after dissipating their accretion discs.  
We note that V819~Tau rotates slower despite being older than V830~Tau;  this suggests that V830~Tau spent more time spinning up since the 
dissipation of its (presumably less massive) accretion disc than V819~Tau and thus that both stars had slightly different accretion histories.  
Although speculative, this scenario would also be consistent with the fact that V819~Tau exhibits traces of leftover dust from the original disc 
whereas V830~Tau shows none \citep[e.g.,][]{Cieza13}.  

Applying our tomographic imaging code \citep[adapted to the case of wTTSs, see][]{Donati14} to our new data set, we derived the surface brightness 
maps and magnetic topologies of both stars.  Cool spots and warm plages are found to be present at the surfaces of both V819~Tau and V830~Tau, but 
with a smaller contrast than the brightness features previously mapped on LkCa~4.  This is especially true for V830~Tau despite the fact that it 
rotates about twice as fast as V819~Tau and was thus expected to have built up, e.g., a large high-contrast polar spot like that of V410~Tau or 
many other similar rapidly-rotating young low-mass stars.  This may suggest that V830~Tau was in a lower-than-usual state of activity at the time of 
our observations\footnote{Although long-term photometric monitoring shows that V830~Tau was indeed exhibiting lower-than-usual photometric 
variability at the time of our observations \citep{Grankin08}, previous studies \citep[e.g.,][]{Skelly10} demonstrated that overall activity of 
low-mass stars, and more specifically the size of their polar spots, are not necessarily correlated with the amplitude of photometric variability.}. 
As for LkCa~4, we find that the brightness maps reconstructed for both stars are in very good agreement with our contemporeneous 
photometric observations.  

The large-scale magnetic fields we reconstruct for V819~Tau and V830~Tau are similar, both in terms of intensities and topologies, 
and are found to be 80--90\% poloidal;  the poloidal component is dominated (especially at some distance from the stars) by a dipolar term of 
polar strength 350--400~G inclined at $\simeq$30\degr\ to the rotation axis whereas the octupolar term is weaker than the dipolar one.  
We stress that neither star shows a conspicuous ring of strong toroidal fields as that reported for LkCa~4 \citep{Donati14}.  
That both stars have similar fields despite their $\times$2 ratio in rotation rates does not come as a surprise given that 
dynamo action is expected to be largely saturated for these active wTTSs.  More unexpected is that these fields are much weaker than that 
of GQ~Lup \citep[typically hosting a 1~kG dipole and a 2~kG octupole, see][]{Donati12} despite their proximity in the HR diagram.  
Although a topological difference (the afore-mentioned toroidal component) was also reported between the large-scale fields of LkCa~4 
on the one hand, and those of AA~Tau and BP~Tau on the other hand (located near LkCa~4 in the HR diagram), their poloidal components 
were similar both in terms of strengths and topologies.  

It is obviously still too early to assess whether magnetic topologies of cTTSs and wTTSs are similar or dissimilar in the regions of the 
HR diagram were both can coexist.  Our observations so far suggest that they differ, either in their poloidal components \citep[this new 
result compared to][]{Donati12} or in their toroidal components \citep[the case of LkCa~4 vs AA~Tau and BP~Tau,][]{Donati14}.  The reason 
for this difference, if confirmed, will need clarification and is potentially complex, acting apparently on both field strengths and topologies.  
Having the potential power to significantly affect atmospheric properties of cTTSs \citep[and thus their location in the HR diagram,][]{Baraffe09} 
and / or to alter their large-scale topologies (by stressing them through disc/star mass transfers and / or by modifying the underlying dynamo 
processes), accretion can participate in this process.  However, the similarity between magnetic topologies of cTTSs and M dwarfs with 
comparable internal structures \citep{Gregory12, Donati14} suggests the opposite.  The temporal variability inherent to dynamo processes 
is another option;  repeated observations of magnetic fields of cTTSs and wTTSs so far \citep[e.g.,][]{Donati11, Donati12, Donati13} however 
indicate that the amplitude of such variations with time is smaller than the one we report here.  Clearly, we will have to wait for more 
MaTYSSE studies in the line of our two first ones to progress on this issue.  

Our data also clearly demonstrate that brightness maps and / or magnetic topologies of V830~Tau and V819~Tau experience very little latitudinal 
shear in a timescale of $\simeq$1~month.  In the particular case of V830~Tau, whose high \vsini\ yields the best spatial resolution at the surface 
of the star, we estimated the amount of differential rotation at a high-enough precision to claim that it is different from zero 
(i.e., not rotating as a solid body) at a confidence level larger than 99.99\%, and smaller than that of the Sun by typically 4.4$\times$ 
(with a time for the equator to lap the pole by one complete rotation cycle equal to $\simeq$500~d, as opposed to $\simeq$110~d for the Sun);  
moreover, the differential rotation estimates we derived from the 
brightness and magnetic maps of V830~Tau, as well as that obtained from the magnetic maps of V819~Tau, all agree together within the error bars.  
Our results also agree with the few existing estimates of differential rotation at the surfaces of similar low-mass wTTSs \citep{Skelly10, Donati14}.  

As for LkCa~4, we finally exploited the results of our tomographic imaging of V819~Tau and V830~Tau to model and filter-out the activity jitter in 
the RV curves of both stars.  For V819~Tau, we find that the activity jitter is filtered down to a rms RV precision of 0.033~\kms, further 
demonstrating that our technique works well and performs as expected from our initial set of simulations \citep{Donati14}.  This leaves little 
hope to find potential hJs around V819~Tau using our technique - despite its exhibiting traces of dust, possibly indicating planet formation, 
which made it an obvious target for our programme.  
For V830~Tau however, the RV residuals after filtering exhibit a significantly larger rms dispersion of 0.104~\kms\ caused by variability on 
timescales of a few days;  we find that this dispersion excess is unlikely to relate to intrinsic (e.g., flaring-like) activity (since RV residuals 
do not correlate with activity proxies) nor to instrumental effects (that would impact data from both stars).  One option is that the observed excess 
dispersion is caused by a $\simeq$1.4~\mjup\ giant planet orbiting V830~Tau in $6.0\pm0.2$~d at a distance of 0.065~AU from the central star.  
Although our detection of a residual RV signal is in principle reliable with a high confidence level (given the excess RV dispersion it generates), 
we nevertheless stress that it relies on a small number of points, preventing us to firmly assess its periodic nature;  as a result,  
we cannot exclude the option that it reflects yet unclear systematics in our data rather than the true presence of a giant planet.  

Additional data on V830~Tau with a denser sampling of both 
rotational and orbital cycles are required to obtain a definite and independent validation of this first promising result.  If confirmed, our 
result would suggest that hJs are likely much more frequent around wTTSs than around MS stars, to ensure that one is detected in a randomly selected 
sample of only a few stars.  The ongoing MaTYSSE observations will enable to further clarify this key issue for our understanding of the formation of 
planetary systems.  Thanks to its much higher sensitivity to wTTSs and enhanced RV precision, SPIRou, the nIR spectropolarimeter / high-precision 
velocimeter currently in construction for CFHT (first light planned in 2017), will tremendously boost this research programme in the near future.

\section*{Acknowledgements} This paper is based on observations obtained at the Canada-France-Hawaii Telescope (CFHT), operated by the National Research 
Council of Canada, the Institut National des Sciences de l'Univers of the Centre National de la Recherche Scientifique (INSU/CNRS) of France and the University 
of Hawaii.  We thank the CFHT QSO team for the its great work and effort at collecting the high-quality MaTYSSE data presented in this paper.  
MaTYSSE is an international collaborative research programme involving experts from more than 10 different countries 
(France, Canada, Brazil, Taiwan, UK, Russia, Chile, USA, Switzerland, Portugal, China and Italy).  
We also warmly thank the IDEX initiative at Universit\'e F\'ed\'erale Toulouse Midi-Pyr\'en\'ees (UFTMiP) for funding the STEPS collaboration program between 
IRAP/OMP and ESO and for allocating a ``Chaire d'Attractivit\'e'' to GAJH allowing her regularly visiting Toulouse to work on MaTYSSE data.  
We acknowledge funding from the LabEx OSUG@2020 that allowed purchasing the ProLine PL230 CCD imaging system installed on the 1.25-m telescope at CrAO.
SGG acknowledges support from the Science \& Technology Facilities Council (STFC) via an Ernest Rutherford Fellowship [ST/J003255/1].  
SHPA acknowledges financial support from CNPq, CAPES and Fapemig.  AAV acknowledges support from the Swiss National Science Foundation (SNSF) via the 
allocation of an Ambizione Followship.

\bibliography{v830tau}
\bibliographystyle{mnras}

\bsp	
\label{lastpage}
\end{document}